\documentclass[trackchanges,twocolumn]{aastex701}

\newcommand\stname{TIC118798035}
\newcommand\plnameb{TIC118798035\,b}
\newcommand\plnamec{TIC118798035\,c}
\newcommand\plnamed{TIC118798035\,d}
\newcommand{\rPb}{$11.507\ d$}
\newcommand{\rPc}{$22.564\ d$}
\newcommand{\rPd}{$48.925\ d$}

\newcommand{\rjup}{$R_J$}
\newcommand{\mjup}{$M_J$}

\newcommand{\rpb}{$0.655\pm0.018$ \rjup}
\newcommand{\rpc}{$0.973\pm0.023$ \rjup}
\newcommand{\rpd}{$0.923\pm0.044$ \rjup}
\newcommand{\mpb}{$0.0250\pm0.0023$ \mjup}
\newcommand{\mpc}{$0.403\pm0.024$ \mjup}
\newcommand{\mpd}{$0.773\pm0.052$ \mjup}

\begin{document}

\title{A Compact Multi-Planet System of Three Transiting Giant Planets Around TIC118798035}

\author[0000-0002-9158-7315]{Rafael Brahm}
\affil{Facultad de Ingenier\'ia y Ciencias, Universidad Adolfo Ib\'{a}\~{n}ez, Av. Diagonal las Torres 2640, Pe\~{n}alol\'{e}n, Santiago, Chile}
\affil{Millennium Institute for Astrophysics, Nuncio Monse\~{n}or Sotero Sanz 100, Of. 104, Providencia, Santiago, Chile}
\email{rafael.brahm@uai.cl}

\author[0000-0002-0236-775X]{Trifon Trifonov} 
\affiliation{Max-Planck-Institut für Astronomie, Königstuhl 17, D-69117 Heidelberg, Germany}
\affiliation{Department of Astronomy, Sofia University ``St Kliment Ohridski'', 5 James Bourchier Blvd, BG-1164 Sofia, Bulgaria}
\affiliation{Landessternwarte, Zentrum f\"ur Astronomie der Universit\"at Heidelberg, K\"onigstuhl 12, D-69117 Heidelberg, Germany}
\email{trifonov@mpia.de}

\author[0000-0002-5389-3944]{Andr\'es Jord\'an}
\affil{Facultad de Ingenier\'ia y Ciencias, Universidad Adolfo Ib\'{a}\~{n}ez, Av. Diagonal las Torres 2640, Pe\~{n}alol\'{e}n, Santiago, Chile}
\affil{Millennium Institute for Astrophysics, Nuncio Monse\~{n}or Sotero Sanz 100, Of. 104, Providencia, Santiago, Chile}
\email{andres.jordan@uai.cl}

\author[0000-0002-1493-300X]{Thomas Henning} 
\affiliation{Max-Planck-Institut für Astronomie, Königstuhl 17, D-69117 Heidelberg, Germany}
\email{henning@mpia.de}

\author[0000-0001-9513-1449]{N\'estor Espinoza} 
\affiliation{Space Telescope Science Institute, 3700 San Martin Drive, Baltimore, MD 21218, USA}
\email{nespinoza@stsci.edu}

\author[0000-0003-3047-6272]{Felipe I. Rojas} 
\affiliation{Instituto de Astrof\'isica, Pontificia Universidad Cat\'olica de Chile, Av. Vicu\~na Mackenna 4860, 7820436 Macul, Santiago, Chile}
\email{firojas@uc.cl}

\author[0009-0004-8891-4057]{Marcelo Tala Pinto} 
\affil{Facultad de Ingenier\'ia y Ciencias, Universidad Adolfo Ib\'{a}\~{n}ez, Av. Diagonal las Torres 2640, Pe\~{n}alol\'{e}n, Santiago, Chile}
\affil{Millennium Institute for Astrophysics, Nuncio Monse\~{n}or Sotero Sanz 100, Of. 104, Providencia, Santiago, Chile}
\email{marcelo.tala@edu.uai.cl}

\author{Mat\'ias I. Jones} 
\affiliation{European Southern Observatory (ESO), Alonso de C\'ordova 3107, Vitacura, Casilla 19001, Santiago, Chile}
\email{mjones@eso.org}

\author{Daniel Thorngren}
\affiliation{Department of Physics and Astronomy, Johns Hopkins University, Baltimore, MD 21210, USA}
\email{dpthorngren@jhu.edu}

\author[0000-0002-9147-7925]{Lorena Acuña}
\affiliation{Max-Planck-Institut für Astronomie, Königstuhl 17, D-69117 Heidelberg, Germany}
\email{acuna@mpia.de}

\author[0000-0003-3130-2768]{Jan Eberhardt} 
\affiliation{Max-Planck-Institut für Astronomie, Königstuhl 17, D-69117 Heidelberg, Germany}
\email{eberhardt@mpia.de}

\author{Yared Reinarz} 
\affiliation{Max-Planck-Institut für Astronomie, Königstuhl 17, D-69117 Heidelberg, Germany}
\email{reinarz@mpia.de}

\author[0009-0009-8795-4563]{Helem Salinas} %
\affiliation{Facultad de Ingenier\'ia y Ciencias, Universidad Adolfo Ib\'{a}\~{n}ez, Av. Diagonal las Torres 2640, Pe\~{n}alol\'{e}n, Santiago, Chile}
\email{yes.helem@gmail.com}

\author[0000-0002-2994-2929]{Michaela V\'{i}tkov\'{a}}
\affiliation{Astronomical Institute of the Czech Academy of Sciences, Fri\v{c}ova 298, CZ-25165 Ond\v{r}ejov, Czech Republic}
\affiliation{Department of Theoretical Physics and Astrophysics, Faculty of Science, Masaryk University, Kotl\'{a}\v{r}sk\'{a} 2, CZ-61137 Brno, Czech Republic}
\email{michaela.vitkova@asu.cas.cz}

\author[0000-0001-9480-8526]{Juan I. Espinoza-Retamal}
\affiliation{Department of Astrophysical Sciences, Princeton University, 4 Ivy Lane, Princeton, NJ 08544, USA}
\affiliation{Instituto de Astrof\'isica, Pontificia Universidad Cat\'olica de Chile, Av. Vicu\~na Mackenna 4860, 7820436 Macul, Santiago, Chile}
\affiliation{Millennium Institute for Astrophysics, Nuncio Monse\~{n}or Sotero Sanz 100, Of. 104, Providencia, Santiago, Chile}
\email{jiespinozar@uc.cl}

\author[0000-0001-7204-6727
]{Gaspar Bakos} %
\affiliation{Department of Astrophysical Sciences, Princeton University, 4 Ivy Lane, Princeton, NJ 08544, USA}
\email{gbakos@astro.princeton.edu}

\author[0000-0002-8585-4544]{Attila B\'{o}di}
\affiliation{Department of Astrophysical Sciences, Princeton University, 4 Ivy Lane, Princeton, NJ 08544, USA}
\email{abodi@princeton.edu}

\author[0009-0009-2966-7507]{Gavin Boyle}
\affil{El Sauce Observatory --- Obstech, Coquimbo, Chile}
\affil{Cavendish Laboratory, J. J. Thomson Avenue, Cambridge, CB3 0HE, UK}
\email{gavinsboyle@icloud.com}

\author{Zolt\'{a}n Csubry}
\affiliation{Department of Astrophysical Sciences, Princeton University, 4 Ivy Lane, Princeton, NJ 08544, USA}
\email{zcsubry@astro.princeton.edu}

\author[0000-0001-8732-6166]{Joel Hartman} %
\affiliation{Department of Astrophysical Sciences, Princeton University, 4 Ivy Lane, Princeton, NJ 08544, USA}
\email{joel.d.hartman@gmail.com}

\author{Anthony Keyes}
\affiliation{Department of Astrophysical Sciences, Princeton University, 4 Ivy Lane, Princeton, NJ 08544, USA}
\email{anthony.keyes@princeton.edu}

\author[0000-0001-7070-3842]{Vincent Suc} 
\affil{Facultad de Ingenier\'ia y Ciencias, Universidad Adolfo Ib\'{a}\~{n}ez, Av. Diagonal las Torres 2640, Pe\~{n}alol\'{e}n, Santiago, Chile}
\affil{Millennium Institute for Astrophysics, Nuncio Monse\~{n}or Sotero Sanz 100, Of. 104, Providencia, Santiago, Chile}
\affil{El Sauce Observatory --- Obstech, Coquimbo, Chile}
\email{vincent.suc@gmail.com}

\author[0000-0003-4787-2335]{Geert Jan Talens}
\affiliation{Denys Wilkinson Building, Department of Physics, University of Oxford, OX1 3RH, UK}
\email{talensgj@gmail.com}

\begin{abstract}
We report the discovery and characterization of three transiting giant planets in the \stname\ system. The three planets were identified as transiting candidates from data of the TESS mission, and confirmed with ground-based photometric transit observations along with radial velocity variations obtained with FEROS, HARPS and ESPRESSO. The three planets present transit timing variations (TTVs). We performed a N-body orbital fitting to the TTVs and radial velocities finding that \plnameb\ is as warm low-density Neptune with a mass of \mpb, a radius of \rpb, and an orbital period of \rPb; \plnamec\ is a warm Saturn with a mass of \mpc, a radius of \rpc, and an orbital period of \rPc; and \plnamed\ is a warm Jupiter with a mass of \mpd, a radius of \rpd, and an orbital period of \rPd. The bulk metallicities of the three planets don't fully follow the mass-metallicity correlation found for the giant planets of the solar system, which hints at a somewhat different formation history for the planets of the \stname\ system. \stname\ is the only system having more than two transiting planets larger than 0.5 $R_J$ with a precise orbital and physical characterization, amenable for future atmospheric studies.
\end{abstract}



\section{Introduction}
Our solar system consists of a G-type dwarf star surrounded by eight planets in a coplanar configuration. Four of these eight planets are termed giant planets whose radii are larger than 3.8 $R_{\oplus}$ having extended primordial gaseous atmospheres. Thanks to different \textit{insitu} measurements, it has been possible to conclude that these planets are enriched in metals compared to our Sun \citep{neptune,jupiter,saturn}, and that their metallicity increases as the planet mass decreases. These observations are consistent with the core accretion theory of giant planet formation \citep{pollack}.

In the past few decades, we have significantly improved our knowledge of the general structures and architectures of exoplanetary systems that contain giant planets \citep{winn:2015,dawson:2018}. A significant number of giant planets have been found orbiting in extreme proximity to their parent stars (hot Jupiters, $a<0.1\ AU$). Hot Jupiters are preferentially found with no nearby planet companions \citep{hord}, which has been attributed to high-excentricity migration mechanisms for the formation of this population \citep[e.g.,][]{petrovich:2015}. Warm Jupiters are gas giant planets orbiting beyond $0.1$ AU, but inside the water snow-line. This population of giant planets has been found to have a significantly larger fraction of companions than hot Jupiters \citep{huang2016}, indicating that an important number of warm Jupiters experienced a quiescent disk-driven migration. This is also consistent with some warm Jupiters found in multiple systems close to orbital mean motion resonances \citep[e.g.,][]{HD27894, toi-2202,Bozhilov2023,toi-4504, toi-6695}, and with the recently observed preponderance of warm Jupiters in single-star systems having low stellar obliquities \citep{rice2022,8rms}.

Transiting warm Jupiters are key objects for investigating the interior structure of giant exoplanets. These planets are not subject to the mechanism that inflates the radii of hot Jupiters, and therefore standard interior models of planet structure can be used to infer their bulk metallicities \citep{thorngren:2016, gastli,chachan}. 

In some cases, multiple transiting warm gas giants have been found in the same system \citep[e.g.][]{kepler31,toi216, toi2525}. Nevertheless, planetary systems with more than two transiting giant planets are extremely scarce. These systems, while infrequent, are particularly interesting, because they allow us to investigate if the properties observed in our solar system giants are also present in other environments.

In this work we present the discovery and characterization of an extrasolar system composed of a G-type dwarf star (\stname) orbited by three transiting warm giant planets ($R_P > 5\ R_{\oplus}$) in a compact configuration, that was made in the conetxt of the Warm gIaNts with tEss collaboration \citep[WINE,][]{brahm:2019,jordan2020}.

\section{Observations} \label{sec:obs}
\subsection{TESS}
\begin{figure*}[t!]
    \centering
    \includegraphics[width=\textwidth]{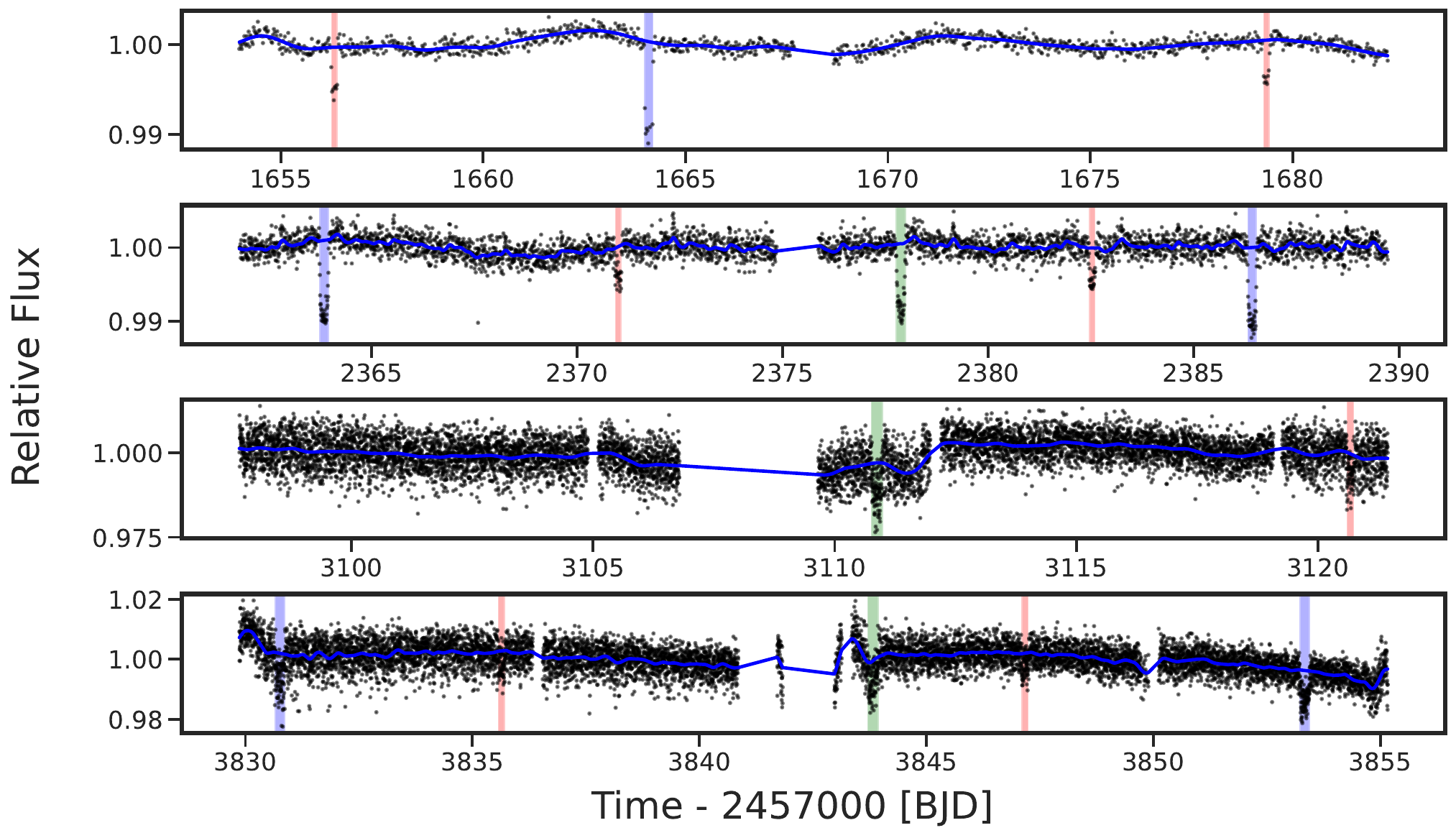}
    \caption{TESS light curves of \stname\ from sectors 13, 39, 66, and 93, from top to bottom. Regions around transits of the candidates $b$, $c$, and $d$ ar marked in red, blue and green, respectively. We also plot the model for the out of transit variations computed with the gaussian process described in Section \ref{sec:juliet}}.
    \label{fig:tess}
\end{figure*}

\stname\ was monitored by the TESS mission \citep{tess} in sectors 13, 39, 66, and 93, with cadences of 1800 s, 600 s, 200 s, and 200 s, respectively. For sectors 13 and 39, we retrieved the SPOC light curves of \stname\  from MAST. For sectors 66 and 93 we generated the light curves using the \texttt{tesseract} pipeline\footnote{\url{https://github.com/astrofelipe/tesseract}}.
These four light curves contain transits of three different planet candidates (b, c, and d) with transit depths that are deep enough to be detected by the eye. \plnameb\ and \plnamec\ were reported as transiting candidates using a machine learning algorithm applied to all TESS-SPOC light curves of the primary TESS mission \citep{salinas:2025}. \plnameb\ has a transit depth of $\approx5000$ ppm and is present in all four TESS sectors. Candidate c has a depth of $\approx9000$ ppm and is present in sectors 13, 39, and 93. Candidate d is reported for the first time in the present study, has transits in sectors 39, 66, and 93, and has a depth of $\approx9000$ ppm.
The rough orbital periods of the two inner candidates are \rPb\ and \rPc, respectively. However, some initial Keplerian fits to these light curves indicated the presence of significant transit timing variations. In the case of \plnamed, given that each TESS sector is separated by $\sim$2 years and that there is only one transit per sector, there were several possible orbital periods for this candidate. The longest possible orbital period for \plnamed\ was of 733.00 d and the shortest one was of 38.58 d. The TESS light curves used in this study are presented in Figure \ref{fig:tess}.

\subsection{Ground-based follow-up observations}

Given that the pixel scale of TESS is relatively large (20\arcsec), seeing limited ground-based observations are required to confirm that the transit signals occur on the target star and not in close fainter neighboring stars.

\subsubsection{Obsrevatoire Moana}

\begin{figure}[t!]
    \centering
    \includegraphics[width=0.9\columnwidth]{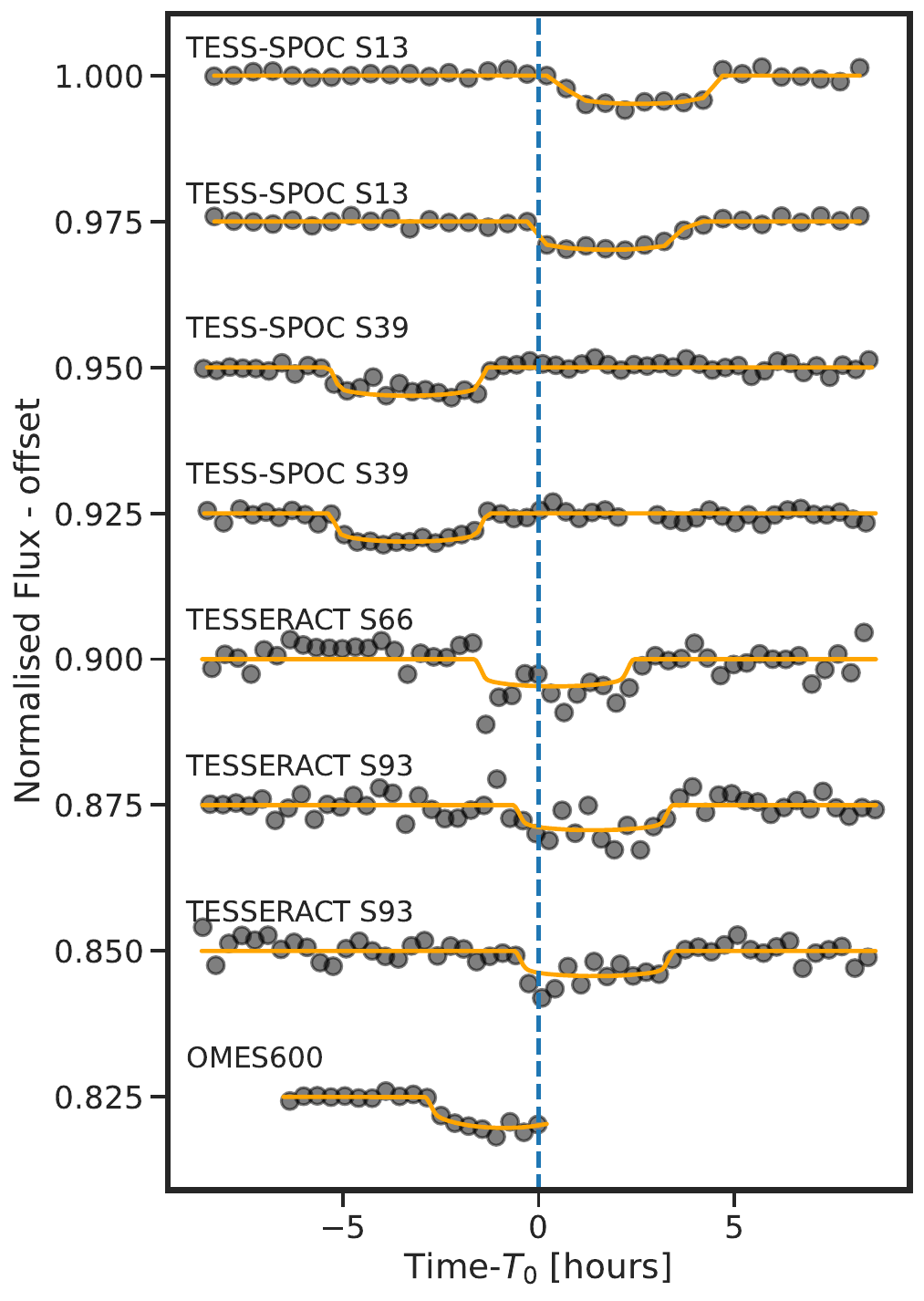}
    \caption{TESS and ground-based transits of \plnameb\ centered on the transit time predicted from linear ephemeris. The transit model generated by  \texttt{juliet} is also plotted for each transit. }
    \label{fig:transitsb}
\end{figure}

\begin{figure}[t!]
    \centering
    \includegraphics[width=0.9\columnwidth]{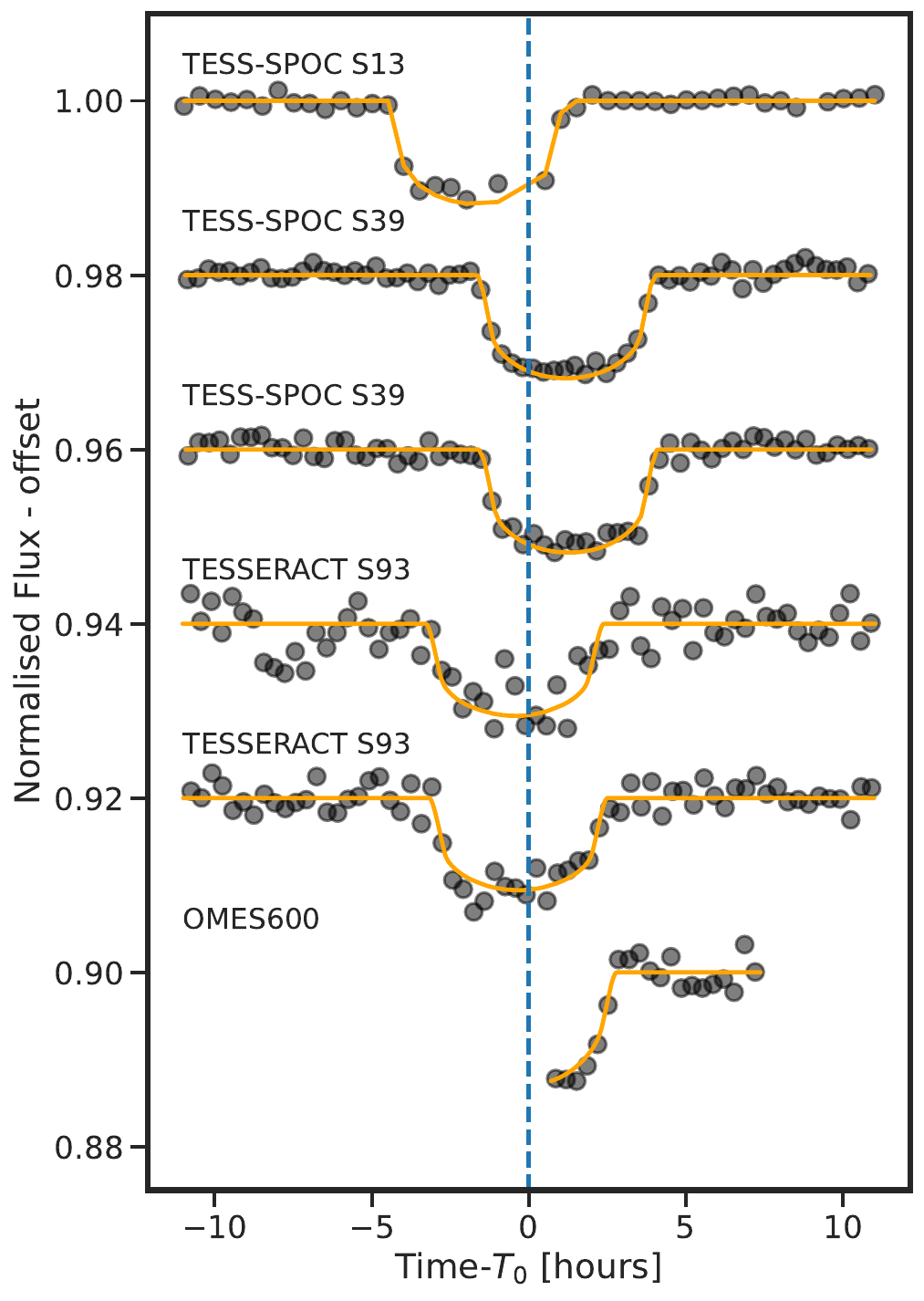}
    \caption{Same as Figure \ref{fig:transitsb}, but for \plnamec.}
    \label{fig:transitsc}
\end{figure}

Observatoire Moana is a network of small-aperture (0.5 m and 0.6 m) robotic telescopes with two stations in El Sauce Observatory in Chile, one station in Siding Spring Observatory in Australia, one station in Spain, and two stations in USA.
We used the 0.6m telescope at El Sauce (OMES600) to monitor one partial transit of \plnameb\ and one partial transit of \plnamec.
We observed \plnamec\ on 10 August 2025, and \plnameb\ on 16 August 2025. In both cases we used the sloan r' filter and exposure times of 29 s.
The data was processed with a dedicated and automated pipeline \citep{toi2525, brahm:2023}. The OMES600 light curves are presented in Figures \ref{fig:transitsb} and \ref{fig:transitsc}.

\subsubsection{HATPI}
HATPI\footnote{\url{https://hatpi.org}} is a photometric instrument installed at Las Campanas Observatory (LCO) of the Carnegie Institution for Science.
HATPI consists of a mosaic array of 64 identical lenses and cameras installed in a single mount, imaging the entire visible sky from LCO at 30 and 45 second cadence and 20\arcsec pixel scale (Bakos et al 2025, in prep).
We retrieved the HATPI light curve of TIC118798035 through the HATPI data portal. These observations extend from 2024-02-27 to 2024-11-04 and were taken during HATPI's ``shakedown operations'' period.

\subsection{High resolution spectroscopy}
High-resolution spectroscopy is required to measure precision radial velocities on the target star to identify periodic signals that can be attributed to the transiting candidates. These observations are fundamental in confirming the planetary nature of the transiting candidates. We used three different spectrographs to monitor the radial velocity variations of \stname. The radial velocity values are presented in Table \ref{tab:rvs}. These radial velocities have significant variations with amplitudes consistent with those produced by gas giant planets.
\subsubsection{FEROS}
\stname\ was observed between July of 2024 and September of 2025 with the FEROS spectrograph \citep{feros} installed in the MPG 2.2 m telescope, at the ESO La Silla Observatory. We obtained 27 spectra using exposure times of 1200 s. The observations were performed using the simultaneous calibration technique to trace instrumental drifts of the wavelength solution. FEROS data were processed with the automated \texttt{ceres} pipeline \citep{ceres} to obtain precision radial velocities and bisector span measurements using the cross-correlation technique. We used the G2 binary mask as a template. We also computed the full width at half maximum (FWHM) of the cross-correlation peak and the S-index that traces the stellar activity. 
\subsubsection{HARPS}
We used the HARPS spectrograph \citep{harps} installed at the ESO 3.6 m telescope, at the ESO La Silla Observatory, to monitor \stname\ between October of 2024 and September of 2025. 16 observations were obtained with exposure times of 1800 s.  HARPS data were reduced with the dedicated HARPS data reduction software, and radial velocities were calculated with \texttt{serval} \citep{serval}, which uses a template matching algorithm. We also independently obtained the bisector span measurements, FWHM and S-index values.
\subsubsection{ESPRESSO}
We also used the ESPRESSO spectrograph \citep{espresso} installed at the VLT, at the ESO Paranal Observatory, to monitor \stname\ between October of 2024 and March of 2025. We obtained 7 spectra using exposure times of 600 s. ESPRESSO data was processed with the Data Reduction Software (DRS) version 3.3.0, and we used the radial velocities computed by the DRS. The DRS also delivers the measurement of the bisector span and FWHM. We obtained the stellar activity S-index from the ESPRESSO DAS Pipeline (version 1.3.8).

\section{Analysis} \label{sec:floats}
\subsection{Stellar Analysis}
\label{stpars}
We adopted the procedure described in \citet{brahm:2019} to characterize the host star. This process is a two-step iterative methodology. The first step is to use the co-added HARPS spectrum to measure the atmospheric stellar parameters ($T_{\mathrm{eff}}$, $\log(g)$, [Fe/H], and $v\sin(i)$) using the \texttt{zaspe} package \citep{zaspe}, which compares the high resolution spectrum against a grid of synthetic models. The second step consists of a spectral energy distribution fit to public broad-band photometry of the star using the PARSEC stellar evolution models \citep{parsec} and the GAIA DR3 \citep{DR3} parallax. In this step we use the spectroscopic derived $T_{\mathrm{eff}}$ as a prior and we fix the metallicity of that obtained with \texttt{zaspe}. With this SED fit we obtain a more precise measurement of the stellar $\log(g)$ which we held fix in a new run of \texttt{zaspe}. The iteration continues until reaching convergence in the measured atmospheric parameters.
The results of our stellar analysis are presented in Table \ref{tab:stellar}. We find that \stname\ is a metal rich ([Fe/H]=$+0.20\pm0.05$) G-type dwarf star not too different from our own Sun.

\subsection{Radial velocities}
\label{sec:rvs}
\begin{figure}[t!]
    \centering
    \includegraphics[width=\columnwidth]{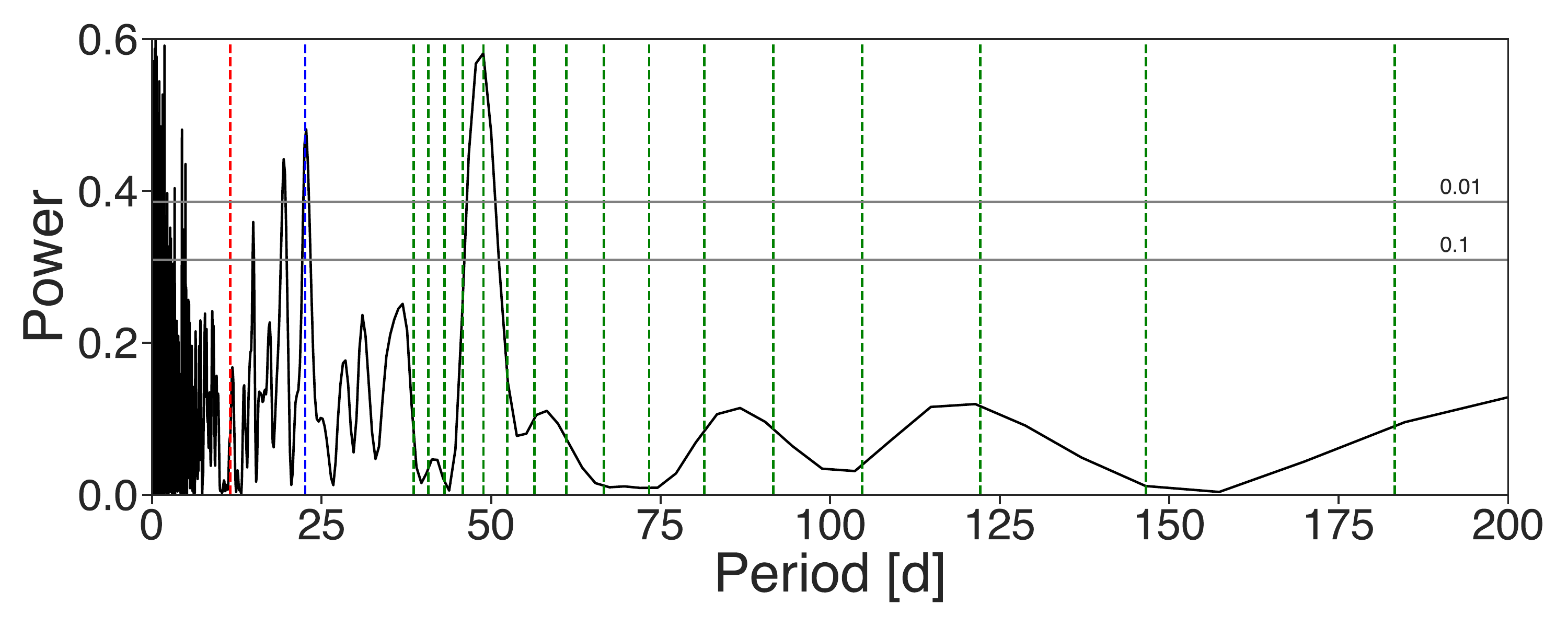}
    \caption{GLS periodogram generated from the FEROS, HARPS, and ESPRESSO radial velocities of \stname. The horizontal lines correspond to the 10\% and 1\% false alarm probabilities. The red and blue dashed vertical lines mark the orbital periods of candidates $b$ and $c$, respectively. The green dashed vertical lines mark the possible period aliases of \plnamed\ based on TESS photometry.}
    \label{fig:gls}
    \centering
    \includegraphics[width=\columnwidth]{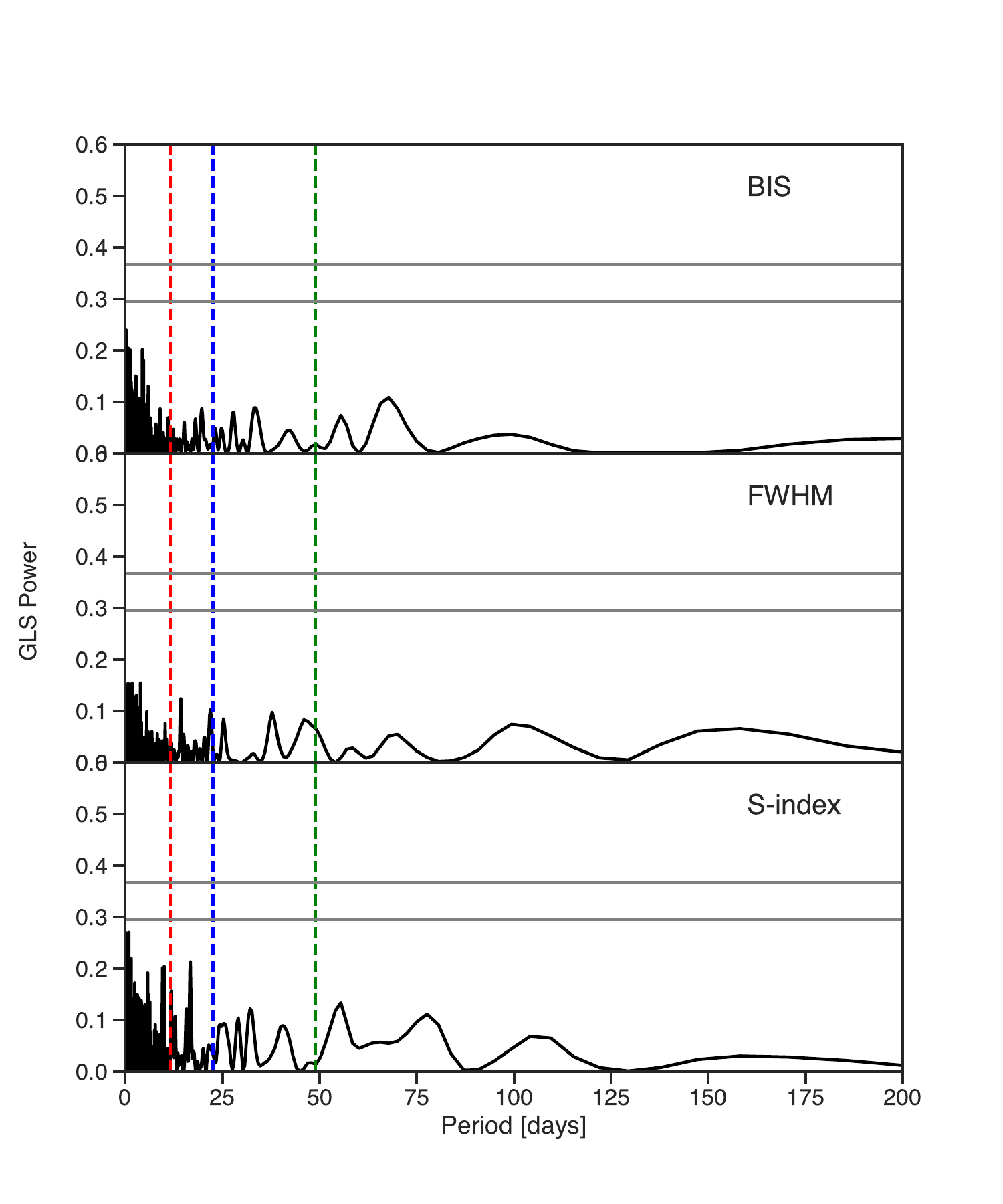}
    \caption{Same as in Figure \ref{fig:gls} but for the bisector spans, FWHM, and S-index. Here the green dashed line corresponds to the orbital period of \plnamed\ constrained by the radial velocities and HATPI light curve.}
    \label{fig:glsbis}
\end{figure}

We computed the Generalized Lomb Scargle (GLS) periodogram after combining the radial velocities of FEROS, HARPS and ESPRESSO in a single data set. We take into account the errors of the radial velocities to generate the GLS. The periodogram is presented in Figure \ref{fig:gls}. It is possible to distinguish two significant peaks above five days. The peak with the second highest power is consistent with the orbital period of \plnamec.  There is no significant peak close to the orbital period of \plnameb, probably due to the relatively low mass of this smaller giant planet. The peak of highest significance on the periodogram is consistent with one of the possible orbital periods of \plnamed\ (48.8 d), which is strong evidence in favor of this period alias for the true period of this outer planet.
We also computed the GLS periodograms for the bisector span, FWHM, and S-index measurements, which are shown in Figure \ref{fig:glsbis}. There are no significant peaks in these periodograms, supporting the view that the signals present in the radial velocities are of planetary origin and not produced by stellar activity or false positive scenarios.

\begin{deluxetable*}{llcr}
\tablecaption{Stellar properties$^a$ of \stname.  \label{tab:stellar}}
\tablecolumns{4}
\tablewidth{0pt}
\tablehead{Parameter & Description & Value & Reference}
\startdata
RA & Right Ascension (J2015.5) & 18h14m54.77s & {\it Gaia} DR3$^b$\\
Dec & Declination (J2015.5) & -54d26m02.6s & {\it Gaia} DR3\\
pm$^{\rm RA}$ & Proper motion in RA (mas/yr) & 19.86$\pm$0.05 & {\it Gaia} DR3\\
pm$^{\rm Dec}$ & Proper motion in DEC (mas/yr) & 6.45$\pm$0.05 & {\it Gaia} DR3\\
$\pi$ & Parallax (mas) & 4.04$\pm$0.02 &  {\it Gaia} DR3 \\
$d$ & Distance (pc) & 242$\pm$1.2 & {\it Gaia} DR3 \\
\hline
T & TESS magnitude (mag) & 11.294$\pm$0.006 & TICv8$^c$\\
B  & B-band magnitude (mag) & 12.7$\pm$0.4 & APASS$^d$\\
V  & V-band magnitude (mag) &  11.99$\pm$0.03 & APASS\\
G  & {\it Gaia} G-band magnitude (mag) & 11.787$\pm$0.002 & {\it Gaia} DR3\\
G$_{\rm BP}$ & {\it Gaia} BP-band magnitude (mag) &  12.190 $\pm$ 0.005  & {\it Gaia} DR3\\
G$_{\rm RP}$ & {\it Gaia} RP-band magnitude (mag) &  11.243 $\pm$ 0.003  & {\it Gaia} DR3\\
J & 2MASS J-band magnitude (mag) & 10.59 $\pm$ 0.02  & 2MASS$^e$\\
H & 2MASS H-band magnitude (mag) & 10.28 $\pm$ 0.02  &  2MASS\\
K$_s$ & 2MASS K$_s$-band magnitude (mag) & 10.21 $\pm$0.02 & 2MASS\\
\hline
$T_{\rm eff}$ & Effective temperature (K) & 5820 $\pm$ 80 & This work\\
$\log{g}$ & Surface gravity (cgs) & 4.43 $\pm$ 0.02    & This work\\
$[$Fe/H$]$ & Metallicity (dex) & +0.20 $\pm$ 0.05  & This work\\
$v\sin{i_\star}$ & Projected rotational velocity (km/s) & 3.2 $\pm$ 0.5   & This work\\
$M_{\star}$ & Mass ($M_\odot$) & 1.02 $\pm$ 0.03 & This work\\
$R_{\star}$ & Radius ($R_\odot$) & 1.02 $\pm$ 0.02  & This work\\
L$_{\star}$ & Luminosity ($L_\odot$) & 0.97 $\pm$ 0.02 & This work\\
A$_{V}$ & Visual extinction (mag) & 0.18 $\pm$ 0.02 & This work\\
Age & Age (Gyr) & 4.4$_{-1.6}^{+1.7} $ & This work\\
$\rho_\star$ & Density (g/cm$^{3}$) & 1.36$_{-0.08}^{+0.09}$ & This work\\
\enddata
\tablecomments{$^a$ The errors in the stellar parameters presented in this study don't consider differences among different stellar evolutionary models \citep{tayar:2022} and can be underestimated.\\
$^b$\citet{DR3}.\\
$^c$ \citet{Stassun2018,Stassun2019}. The TESS magnitude is shown only for reference and was not included in our stellar analysis.\\
$^d$ \citet{apass}.\\
$^e$ \citet{2mass}.}
\end{deluxetable*}

\subsection{Light Curve Analysis}
\label{sec:lcs}
\subsubsection{Transit of \plnamed\ in HATPI}
\label{sec:mhatpi}
\begin{figure}[t!]
    \centering
    \includegraphics[width=0.9\columnwidth]{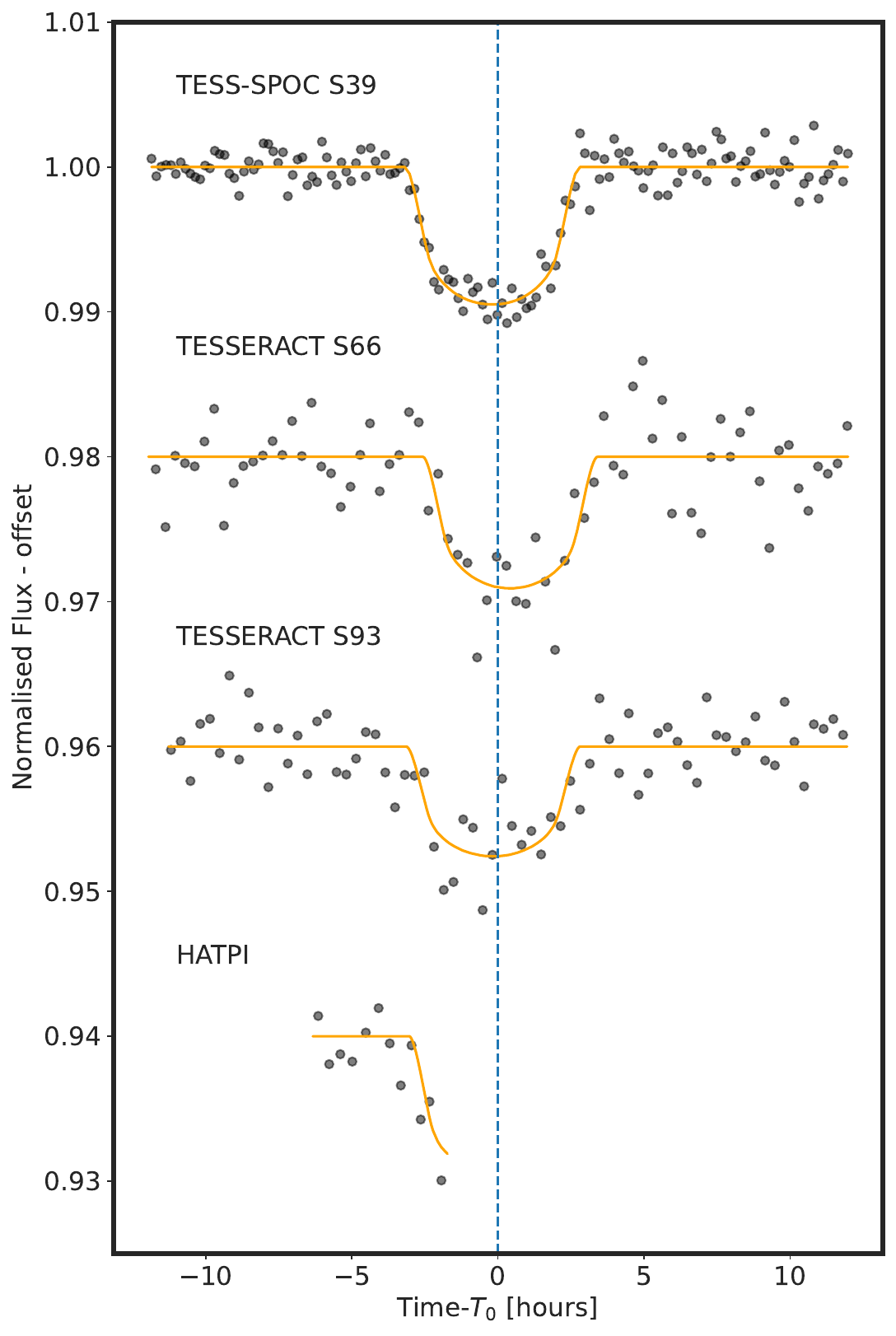}
    \caption{Same as Figure \ref{fig:transitsb}, but for \plnamed.}
    \label{fig:transitsd}
\end{figure}

The radial velocity analysis presented in Section \ref{sec:rvs} allowed us to constrain the most probable orbital period for \plnamed\ to 48.87 d. We inspected the HATPI light curve and identified that during the night of 10 July 2024 an ingress-like feature is present in the data, whose timing is consistent with the \rPd\ orbital period. This event rules out all other possible period aliases and confirms that the orbital period of \plnamed\ is of \rPd. The HATPI light curve for this night is presented in Figure \ref{fig:transitsd}.

We also used the full HATPI light curve to infer the rotational period of the star. We computed the GLS of the light curve and found a significant peak at $P_{rot}=16.4\pm 1.6$ days. If we combine this value with the radius of the star derived from our stellar analysis we obtain an equatorial rotational velocity of $v_{eq}=3.1\pm0.3\ km/s$. This value is fully consistent with the projected rotational velocity obtained with \texttt{zaspe}, which implies an inclination of the stellar rotation axis close to 90 deg with respect to the line of sight, and a probable low stellar obliquity for the \stname\ planetary system.

\subsubsection{Light curve modeling and TTV extraction}
\label{sec:juliet}
We used the TESS light curves of the four sectors along with the OMES600 and HATPI light curves to model the transits of the three planets present in the \stname\ system.
We first masked out all the transits present in the TESS light curves and used the \texttt{juliet} package \citep{juliet} to perform a fit to the out of transit flux in order to de-trend these light curves. The light curve of each TESS sector was modeled with a different gaussian process. In all cases we used  a Matern 3/2 Kernel.

We then used the detrended TESS light curves and the ground-based light curves to model the transits of the \stname\ system with \texttt{juliet}. We adopted the parametrization that allows the user to fit the times of each transit individually instead of fitting the orbital period and time of the first transit. We applied this parametrization given the presence of significant TTVs for the three planets. For this analysis, we assumed circular orbits for the three planets. For the TESS light curves we used the quadratic limb darkening law, while for the ground-based light curves we used a linear one. We adopted the parametrization in which we fit for the stellar density instead of the normalized semi-major axis, and used the value obtained in Section \ref{stpars} as a prior. For sectors 66 and 93 of TESS we allowed the dilution factor to vary between 0.5 and 1.0. Given that the SPOC light curves are already corrected by dilution from nearby sources, we fixed the dilution factor to 1. For the ground-based light curves, we assumed no flux contamination by neighbor stars and fixed the dilution factor to 1. We run different \texttt{juliet} fits for the different planets. The results of these three fits are presented in Tables \ref{tab:julietb}, \ref{tab:julietc}, and \ref{tab:julietd}.

\begin{table}[]
\caption{Priors and posteriors for the TTV extraction with {\textsc juliet} for \plnameb.}
\centering
\label{tab:julietb}
\begin{tabular}{ccc}
 \hline 
  \hline
Parameter       & Prior                         & Posterior                           \\ \hline
$b$               & $\mathcal{U}$ (0.0,1.0)       & $0.39^{+0.05}_{-0.07}$           \\
$R_P/R_{\star}$               & $\mathcal{U}$ (0.0,1.0)       & $0.066^{+0.001}_{-0.001}$           \\
$q_{1,TESS}$        & $\mathcal{U}$ (0.0,1.0)       & $0.2^{+0.1}_{-0.1}$           \\
$q_{2,TESS}$        & $\mathcal{U}$ (0.0,1.0)       & $0.5^{+0.3}_{-0.3}$           \\
$q_{1,OMES600}$        & $\mathcal{U}$ (0.0,1.0)       & $0.7^{+0.2}_{-0.2}$           \\
$\rho_{\star}$ [kg cm$^{-3}$]             & $\mathcal{N}$ (1360,100)      & 1390$^{+100}_{-90}$                \\
$m_{\rm dilution,S66}$ & $\mathcal{U}$ (0.5,1.0)                    &  $0.96^{+0.03}_{-0.05}$                               \\
$m_{\rm dilution,S93}$ & $\mathcal{U}$ (0.5,1.0)                    & $0.89^{+0.07}_{-0.08}$                                  \\
$\sigma_{w,S13}$ [ppm]  & $\mathcal{J}$ (0.1,1000.0)    & $4^{+27}_{-3}$      \\
$\sigma_{w,S39}$ [ppm]  & $\mathcal{J}$ (0.1,1000.0)    & $70^{+240}_{-70}$      \\
$\sigma_{w,S66}$ [ppm]  & $\mathcal{J}$ (0.1,1000.0)    & $3880^{+30}_{-30}$      \\
$\sigma_{w,S93}$ [ppm]  & $\mathcal{J}$ (0.1,1000.0)    & $3520^{+30}_{-30}$      \\
$\sigma_{w,OMES600}$ [ppm]  & $\mathcal{J}$ (0.1,1000.0)    & $1300^{+200}_{-200}$      \\
$T_{b,0}$ [BJD] & $\mathcal{N}$ (2458656.3,0.5) & $2458656.321^{+0.002}_{-0.002}$  \\
$T_{b,2}$ [BJD]  & $\mathcal{N}$ (2458679.4,0.5) & $2458679.352^{+0.002}_{-0.002}$  \\
$T_{b,62}$ [BJD] & $\mathcal{N}$ (2459371.0,0.5) & $2459371.000^{+0.002}_{-0.002}$  \\
$T_{b,63}$ [BJD] & $\mathcal{N}$ (2459382.5,0.5) & $2459382.532^{+0.002}_{-0.002}$  \\
$T_{b,127}$ [BJD] & $\mathcal{N}$ (2460120.7,0.5) & $2460120.673^{+0.004}_{-0.004}$  \\
$T_{b,189}$ [BJD] & $\mathcal{N}$ (2460835.7,0.5) & $2460835.638^{+0.008}_{-0.007}$  \\
$T_{b,190}$ [BJD] & $\mathcal{N}$ (2460847.2,0.5) & $2460847.170^{+0.007}_{-0.006}$  \\
$T_{b,195}$ [BJD]& $\mathcal{N}$ (2460904.8,0.5) & $2460904.730^{+0.003}_{-0.003}$  \\
\hline
\hline
\end{tabular}
\end{table}

\begin{table}[]
\caption{Priors and posteriors for the TTV extraction with {\textsc juliet} for \plnamec.}
\centering
\label{tab:julietc}
\begin{tabular}{ccc}
 \hline 
  \hline
Parameter       & Prior                         & Posterior                           \\ \hline
$b$               & $\mathcal{U}$ (0.0,1.0)       & $0.1^{+0.1}_{-0.1}$           \\
$R_P/R_{\star}$               & $\mathcal{U}$ (0.0,1.0)       & $0.098^{+0.001}_{-0.001}$           \\
$q_{1,TESS}$        & $\mathcal{U}$ (0.0,1.0)       & $0.36^{+0.06}_{-0.07}$           \\
$q_{2,TESS}$        & $\mathcal{U}$ (0.0,1.0)       & $0.5^{+0.3}_{-0.3}$           \\
$q_{1,OMES600}$        & $\mathcal{U}$ (0.0,1.0)       & $0.8^{+0.1}_{-0.1}$           \\
$\rho_{\star}$ [kg cm$^{-3}$]             & $\mathcal{N}$ (1360,100)      & 1350$^{+50}_{-70}$                \\
$m_{\rm dilution,S93}$ & $\mathcal{U}$ (0.5,1.0)                    & $0.89^{+0.04}_{-0.04}$                                  \\
$\sigma_{w,S13}$ [ppm]  & $\mathcal{J}$ (0.1,1000.0)    & $3^{+39}_{-3}$      \\
$\sigma_{w,S39}$ [ppm]  & $\mathcal{J}$ (0.1,1000.0)    & $20^{+220}_{-20}$      \\
$\sigma_{w,S93}$ [ppm]  & $\mathcal{J}$ (0.1,1000.0)    & $3500^{+30}_{-30}$      \\
$\sigma_{w,OMES600}$ [ppm]  & $\mathcal{J}$ (0.1,1000.0)    & $3400^{+200}_{-200}$      \\
$T_{c,0}$ [BJD] & $\mathcal{N}$ (2458664.1,0.5) & $2458664.092^{+0.001}_{-0.001}$  \\
$T_{c,31}$ [BJD]  & $\mathcal{N}$ (2459363.9,0.5) & $2459363.847^{+0.001}_{-0.001}$  \\
$T_{c,32}$ [BJD] & $\mathcal{N}$ (2459386.4,0.5) & $2459386.419^{+0.001}_{-0.001}$  \\
$T_{c,96}$ [BJD] & $\mathcal{N}$ (2460830.8,0.5) & $2460830.760^{+0.005}_{-0.006}$  \\
$T_{c,97}$ [BJD] & $\mathcal{N}$ (2460853.3,0.5) & $2460853.334^{+0.003}_{-0.003}$  \\
$T_{c,99}$ [BJD] & $\mathcal{N}$ (2460898.5,0.5) & $2460898.484^{+0.002}_{-0.002}$  \\
\hline
\hline
\end{tabular}
\end{table}

\begin{table}[]
\caption{Priors and posteriors for the TTV extraction with {\textsc juliet} for \plnamed.}
\centering
\label{tab:julietd}
\begin{tabular}{ccc}
 \hline 
  \hline
Parameter       & Prior                         & Posterior                           \\ \hline
$b$               & $\mathcal{U}$ (0.0,1.0)       & $0.65^{+0.04}_{-0.04}$           \\
$R_P/R_{\star}$               & $\mathcal{U}$ (0.0,1.0)       & $0.093^{+0.003}_{-0.003}$           \\
$q_{1,TESS}$        & $\mathcal{U}$ (0.0,1.0)       & $0.6^{+0.3}_{-0.3}$           \\
$q_{2,TESS}$        & $\mathcal{U}$ (0.0,1.0)       & $0.4^{+0.3}_{-0.3}$           \\
$q_{1,HATPI}$        & $\mathcal{U}$ (0.0,1.0)       & $0.7^{+0.2}_{-0.4}$           \\
$\rho_{\star}$ [kg cm$^{-3}$]             & $\mathcal{N}$ (1360,100)      & 1340$^{+100}_{-100}$                \\
$m_{\rm dilution,S66}$ & $\mathcal{U}$ (0.5,1.0)                    & $0.95^{+0.03}_{-0.03}$                                  \\
$m_{\rm dilution,S93}$ & $\mathcal{U}$ (0.5,1.0)                    & $0.80^{+0.09}_{-0.06}$                                  \\
$\sigma_{w,S39}$ [ppm]  & $\mathcal{J}$ (0.1,1000.0)    & $7^{+86}_{-7}$      \\
$\sigma_{w,S66}$ [ppm]  & $\mathcal{J}$ (0.1,1000.0)    & $3880^{+30}_{-30}$      \\
$\sigma_{w,S93}$ [ppm]  & $\mathcal{J}$ (0.1,1000.0)    & $3520^{+30}_{-30}$      \\
$\sigma_{w,HATPI}$ [ppm]  & $\mathcal{J}$ (0.1,1000.0)    & $11^{+203}_{-11}$      \\
$T_{d,0}$ [BJD] & $\mathcal{N}$ (2459377.9,0.5) & $2459377.876^{+0.002}_{-0.002}$  \\
$T_{d,15}$ [BJD]  & $\mathcal{N}$ (2460110.9,0.5) & $2460110.872^{+0.004}_{-0.004}$  \\
$T_{d,23}$ [BJD] & $\mathcal{N}$ (2460501.8,0.5) & $2460501.768^{+0.008}_{-0.009}$  \\
$T_{d,30}$ [BJD] & $\mathcal{N}$ (2460843.9,0.5) & $2460843.817^{+0.007}_{-0.006}$  \\
\hline
\hline
\end{tabular}
\end{table}

\subsection{N-body orbital fitting}

\subsubsection{Methods and Parameterization}

We performed a joint TTV and RV orbital analysis using the {\tt Exo-Striker} fitting tool \citep{Trifonov2019es}, 
which incorporates an internally self-consistent RV N-body model, together with an N-body model for the TTVs based on the {\tt ttvfast} tool. 
A similar approach has already been used in several of our WINE-TTV exoplanet studies, for example, for the TOI-2202 system \citep{toi-2202}, the TOI-2525 system \citep{toi2525}, and the 
TOI-4504 system \citep{toi-4504}. We refer the reader to \citet{toi-2202} and \citet{toi-4504} for further details on the numerical methods 
used for the TTV+RV modeling scheme.


Briefly, our scheme combines three sets of TTVs for each transiting planet (TIC\,118798035\,b, c, and d) and precise RV data from FEROS, HARPS, and ESPRESSO.
The planetary parameterization is done in Jacobi frame  \citep{Lee2003} and includes, the osculating planetary orbital periods $P_{\rm b,c,d}$, and the parameters;
$h_{\rm b,c,d} = e_{\rm b,c,d}\sin\omega_{\rm b,c,d}$,
$k_{\rm b,c,d} = e_{\rm b,c,d}\cos\omega_{\rm b,c,d}$, and
mean longitude angles 
$\lambda_{\rm b,c,d} = \omega_{\rm b,c,d} + M_{\rm b,c,d}$ \citep[see,][]{Tan2013}. 
From these combination of parameters, $e_{\rm b,c,d}$ are the planetary eccentricities, $\omega_{\rm b,c,d}$ are arguments of periastron, and $M_{\rm b,c,d}$, are the mean anomalies.
The planetary dynamical masses in the N-body model were estimated by fitting the radial velocity semi-amplitudes $K_{\rm b,c,d}$.
We allow the planetary inclinations $i_{\rm b,c,d}$ to vary within the derived posterior ranges obtained from the TTV extraction fit. The 
line of node angles $\Omega_{\rm b,c,d}$ needed to characterize the three-dimensional geometry of the orbits fully, however, we keep fixed 
at 0$^\circ$, since these are very difficult to constrain and remain undefined. Therefore, we assumed an edge-on, yet mutually inclined 
system, which is the most likely configuration if we observe the three objects in transit \citep{Ragozzine2010, Brakensiek2016}.
The RVs were modeled with individual instrumental offsets and jitter terms, RV$_{\rm off}$ and RV$_{\rm jitt}$, for the FEROS, HARPS, and ESPRESSO datasets, 
respectively. The osculating orbital parameters are given for the reference epoch ${\rm BJD} = 2458656.3$, selected to precede the first 
observed transit of TIC\,118798035\,b. The N-body model adopts the stellar mass estimate derived from our analysis (see 
Table~\ref{tab:stellar}), namely $M_\star = 1.024 \pm 0.05 M_\odot$. The integration time step of the N-body model was set to $dt = 
0.1$~days, which is very small given the long temporal baseline of the observations, but strictly necessary due to the 
short orbital period of the innermost planet, ensuring sufficient numerical 
accuracy with approximately 100 integration steps per completed orbit.

\subsubsection{Best-fit and posteriors}

To explore the vast orbital parameter space of the three interacting warm giant planets, we performed a global parameter search using the {\tt dynesty} nested sampler \citep[NS;][]{dynesty}, implemented within the {\tt Exo-Striker} tool \citep[see][for details]{toi-4504}. The prior ranges were informed by individual RV analyses, including GLS periodograms, pre-whitening RV cascade fits with multi-Keplerian and N-body models, as well as by the TTV extraction parameters derived from the transit data. We adopted 100 live points per fitted parameter and employed the static NS configuration.

The NS posteriors were calculated primarily to identify the most likely regions of parameter space consistent with the data. Given the high computational cost of the full N-body modelling and the strong parameter degeneracies, a fully converged NS posterior characterisation for 
TIC\,118798035 would require a very large number of NS live-points and an unfeasible number of CPU hours to complete the task. For similar computational constraints, we decided against performing a complete photodynamical analysis of the light curves and RVs, as it would demand excessive 
computational resources with minimal gain compared to the simpler, yet accurate TTV+RV modelling scheme.

Instead, we identified the best $-\ln\mathcal{L}$ solution from the NS run and refined it using an iterative Simplex optimisation \citep{NelderMead} 
to obtain the global best-fit parameters. The posterior probability distributions and the corresponding $1\sigma$ parameter 
uncertainties were subsequently derived with an affine-invariant ensemble Markov Chain Monte Carlo (MCMC) sampler \citep{Goodman2010}, 
implemented through the {\tt emcee} package \citep{emcee}. 

\begin{figure*}
  \centering
  \includegraphics[width=5.85cm]{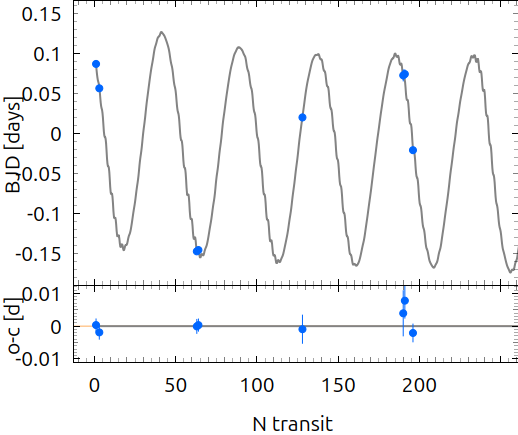}
  \includegraphics[width=5.85cm]{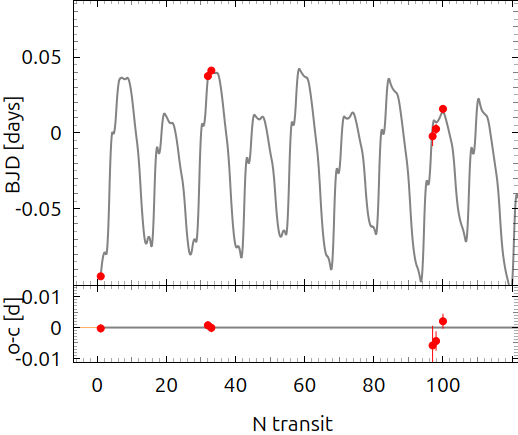} 
  \includegraphics[width=5.85cm]{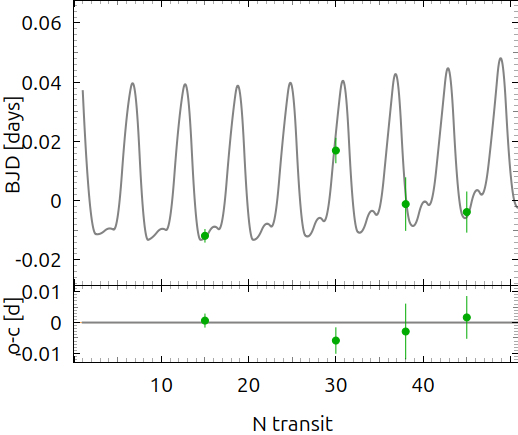}\\

  \caption{TTV time series and best-fit model of TIC 118798035\,b,c and d. The bottom panels show the TTV model residuals.}
  \label{TTVs}
\end{figure*}

\begin{figure*}
  \centering
  \includegraphics[width=5.85cm]{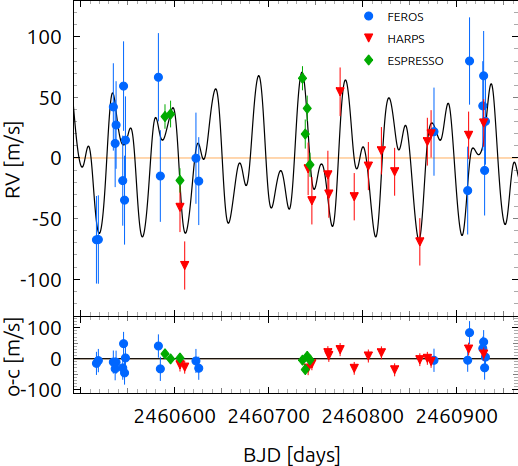}
  \includegraphics[width=5.85cm]{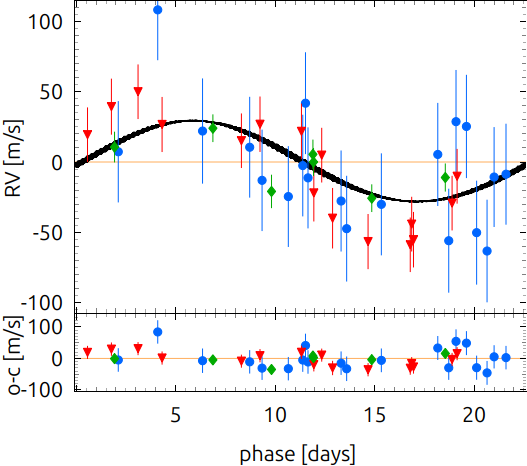} 
  \includegraphics[width=5.85cm]{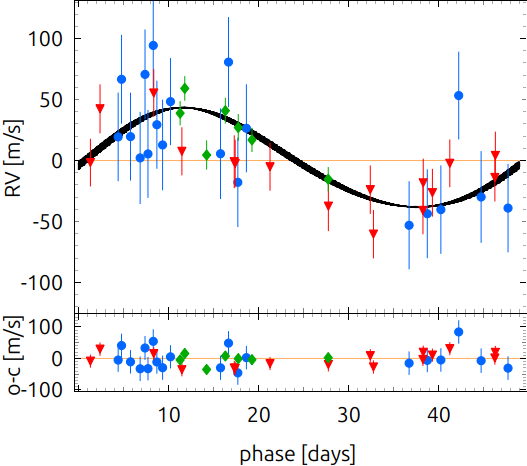}

  \caption{Same model as in \autoref{TTVs}, but applied in the FEROS, HARPS and ESPRESSO, RV time series. 
  The left panel shows the full model to the RVs, the middle panel shows the phased signal for TIC\,118798035\,c, and the right plot shows phased model signal TIC\,118798035\,d. 
  The signal of TIC\,118798035\,b is too small and practically unconstrained, thus is not shown. The bottom panels show the RV model residuals.}
  \label{RVs}
\end{figure*}

\subsubsection{Results}
\label{results}

\noindent

Our MCMC posterior results from the joint TTV+RV modelling of the TIC\,118798035 system are summarised in \autoref{tab:ttv_rv_posterior}, and the corresponding posterior probability distributions are shown in \autoref{mcmc_samp}. In \autoref{tab:ttv_rv_posterior}, we report the posterior median values and their 1$\sigma$ credible intervals derived from the combined TTV and RV data. \autoref{TTVs} and \autoref{RVs} present the resulting best-fit model from our joint N-body TTV+RV analysis, which provides an excellent representation of the observations and is therefore considered robust.

Overall, our analysis reveals that \stname\ hosts a dynamically compact and well-aligned system of three warm giant planets with well-constrained orbital and physical parameters. The joint fit yields a system composed of three gravitationally interacting planets, designated as \plnameb, c, and d, with osculating orbital periods of
$P_{b} = 11.5075_{-0.0015}^{+0.0017}$ days,
$P_{c} = 22.5644_{-0.0024}^{+0.0027}$ days, and
$P_{d} = 48.9244_{-0.0054}^{+0.0048}$ days,
corresponding to semi-major axes of
$a_{b} = 0.10055_{-0.00166}^{+0.00161}$~AU,
$a_{c} = 0.15754_{-0.00260}^{+0.00253}$~AU, and
$a_{d} = 0.26397_{-0.00436}^{+0.00423}$~AU, respectively.

The orbital eccentricities are small but significantly different from circular, with
$e_{b} = 0.0517_{-0.0039}^{+0.0038}$,
$e_{c} = 0.0068_{-0.0055}^{+0.0051}$, and
$e_{d} = 0.0492_{-0.0120}^{+0.0121}$.
Assuming nearly coplanar configurations with small mutual inclinations, and accounting for the stellar mass and its uncertainty, we derive well-constrained dynamical planetary masses of
$m_{b} = 0.0250_{-0.0022}^{+0.0024},M_{\rm J}$,
$m_{c} = 0.403_{-0.025}^{+0.024},M_{\rm J}$, and
$m_{d} = 0.773_{-0.052}^{+0.051},M_{\rm J}$,
corresponding to a sub-Neptune, a Saturn-mass, and a Jupiter-mass planet, respectively.

The three planets are close to first-order commensurability with average period ratios of $P_{c}/P_{b} \approx 1.96$ and $P_{d}/P_{c} \approx 2.17$. However, a detailed N-body inspection of the orbital dynamics does not reveal any evidence of a low-order mean-motion resonance (MMR) between the warm giants.

\noindent
We further performed a comprehensive $N$-body dynamical analysis of the MCMC posterior samples using the symplectic $N$-body integrator {\tt SyMBA} \citep{Duncan1998}, following the same stability setup as applied in previous WINE-TTV studies \citep[see][]{toi-2202,toi2525,toi-6695,toi-4504}. Long-term stability analysis shows that the system is dynamically stable over secular timescales. From 1000 randomly selected MCMC posterior samples integrated over 1~Myr, the TIC\,118798035 planetary system remained stable and dynamically well separated in all cases. Interestingly, about half of the integrated configurations exhibit libration of the secular apsidal angle of planets TIC\,118798035 c and d, defined as $\Delta\omega_{c-d} = \omega_{c} - \omega_{d}$.
The libration occurs around $0^\circ$ with an amplitude of about $70^\circ$, suggesting that the secular coupling between the outer Jovian planets dominates the long-term dynamical evolution of the system.  

\begin{figure*}[!ht]
    \includegraphics[width=5.85cm]{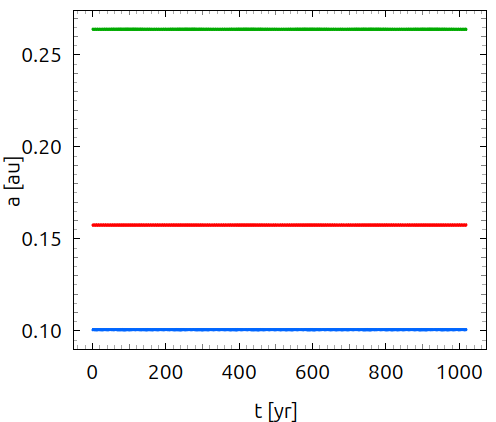} 
    \includegraphics[width=5.85cm]{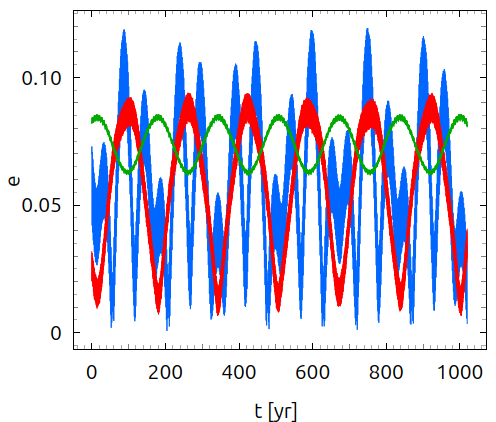} 
    \includegraphics[width=5.85cm]{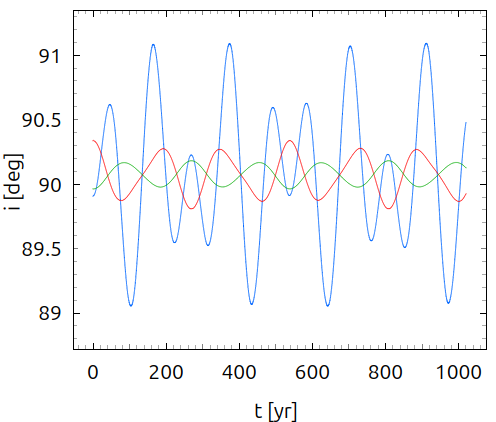} \\
    \includegraphics[width=5.85cm]{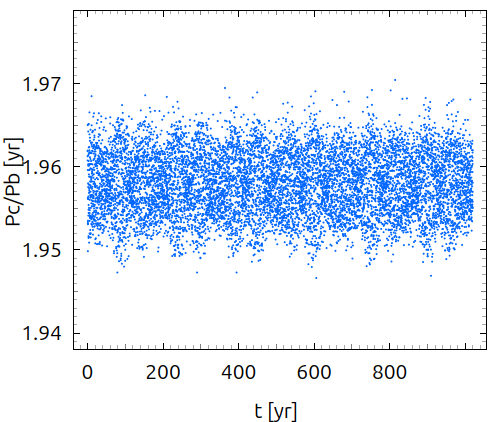} 
    \includegraphics[width=5.85cm]{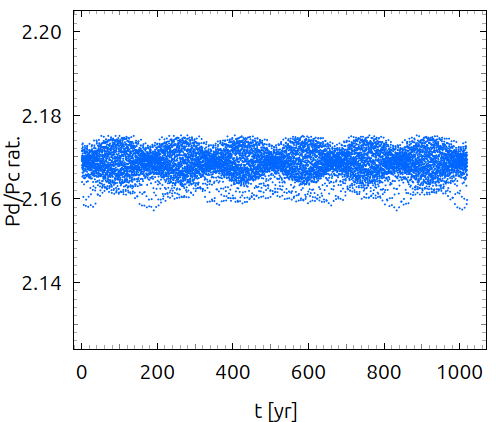} 
    \includegraphics[width=5.85cm]{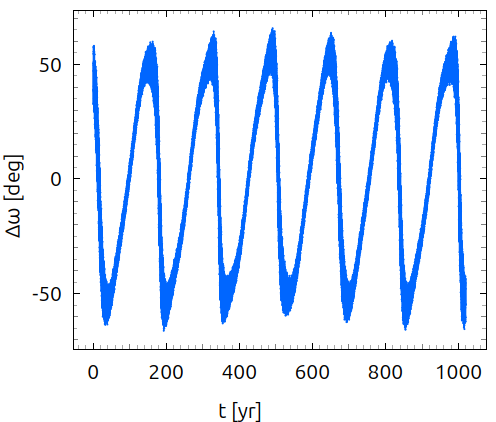} \\
    \caption{Orbital evolution of the TIC\,118798035 system for an extent of 1000 years for the best-fit (maximum $-lnL$) sample from the MCMC posteriors. 
    \emph{Top panels, from left to right:} evolution of the planetary semi-major axes, eccentricities and inclinations where \plnameb\ is shown with blue 
    \plnamec\ with red, and \plnamed\ with green, respectively. 
    \emph{Bottom panels, from left to right} show the evolution of the period ratio between the inner pair $P_{\rm{c}} / P_{\rm{b}}$, the outer pair $P_{\rm{d}} / P_{\rm{c}}$, and the libration of the apsidal angle $\Delta\omega_{c-d}$ in aligned configuration around 0$^\circ$.
    \label{evol_plot}}
\end{figure*}

\autoref{evol_plot} illustrates a representative dynamical evolution of the three-planet system for an extent of 1\,000 yr. In the top panels of \autoref{evol_plot}, from left to right, show the evolution of the semi-major axes, the orbital eccentricities, and the inclinations, respectively. In the bottom panels of \autoref{evol_plot}, from left to right, show the evolution of the period ratio between planets TIC\,118798035 b and c, $P_{\rm{c}}/P_{\rm{b}}$, the period ratio between planets TIC\,118798035 c and d, $P_{\rm{d}}/P_{\rm{c}}$,
and the secular apsidal angle  $\Delta\omega_{\rm{c-d}}$, which clearly exhibits libration around $0^\circ$, consistent with a secular coupling.  

The remaining stable configurations consistent with the TTV and RV data do not show any secular or resonant angle libration, indicating that the system may be chaotic but still long-term stable. It is plausible that the three planets were originally trapped in a Laplace 1:2:4 MMR resonance chain and subsequently escaped from this synchronization due to overstability effects \citep{Goldreich2014}. In this scenario, convergent planetary migration with strongly damped eccentricities could have led only to a transient resonance capture at the 2:1/2:1 commensurability, followed by dynamical relaxation into the present near-resonant configuration.

\subsection{Interior modeling}
To determine the bulk composition of the TIC 118798035 system planets, we use the interior structure models of \citet{thorngren:2019}.  These solve the equations of hydrostatic equilibrium, mass conservation, and an appropriate equation of state to determine the size of a given model planet.  We used the \citet{Thompson1990} equations of state for the heavy elements, which were set at a 50-50 mixture of rock and ice; for hydrogen and helium we used \citet{Chabrier2021} \citep[which incorporates entropy results from][]{Militzer2013}.  To thermally evolve the planets, we use the atmosphere models of \citet{Fortney2007}.

The models are fit to the observed mass, radius, instellation, and age using a Bayesian framework \citep{thorngren:2019} to determine the bulk metallicity.  The model parameters for c and d are the mass, metallicity, and age, with priors for mass and age set to the observed values with uncertainties, and the prior for metallicity is from the \citet{thorngren:2016} mass-metallicity relation (accounting for the wide population spread).

\plnamec\ and d are warm gas giants of fairly typical composition.  For c we obtain a bulk metallicity of $Z= 0.15\pm0.03$, which sets a limit on the atmospheric metallicity \citep{thorngren:2019} of $<19.2\times$Solar.  Relative to the parent star, we have $Z_p/Z_* = 6.72^{+1.6}_{-1.5}$.  Similarly for planet d we have $Z=0.22\pm0.05$, an atmospheric metallicity limit of $<33.2\times$Solar, and $Z_p/Z_* = 9.86^{+2.4}_{-2.6}$.  Note that these uncertainties are largely driven by the shared stellar uncertainties.  As such, the relative metallicity between c and d is better constrained than the absolute metallicity.  For example, if the true stellar radius were 1$\sigma$ above the observational mean, it would imply both planets are larger and therefore have less metal, but the effect would be similar between planets.  Modeling uncertainties (e.g. EOS errors) are not accounted for in the error bars presented, but would also be similar between planets.  Thus the result that d is more metal-rich than c is secure.

Although it is also in the warm giant regime, TIC 118798035 b has a much more ambiguous composition.  As it is a sub-Neptune, we have adjusted our model accordingly, allowing the ice-to-rock ratio and the fraction of ice mixed into the envelope (vs an outer layer of the core) to vary as additional free parameters.  The prior for both additional parameters are set as uniform between 0 and 1.  This covers scenarios in which the planet might be water rich due to forming outside the ice line vs water poor due to forming with in it.  For more massive planets this is a small effect, but for sub-Neptunes it is important.

In this case, for a water-poor model the metallicity is $Z\approx0.6$, whereas more water-rich models allow metallicities exceeding $Z>0.9$.  As such we cannot place a limit on the atmospheric metallicity.  We can only set a lower limit on the bulk metallicity, $Z>0.55$, and relative to the parent star $Z_p/Z_*>20$, but it could easily be much higher.

\begin{table}[htbp]
\centering
\caption{Best-fit (maximum likelihood) parameters and median values with $1\sigma$ credible intervals from the TTV + RV model posterior.}
\begin{tabular}{lcc}
\hline\hline
Parameter & Best-fit (max. lnL) & Median and $1\sigma$ \\
\hline
\multicolumn{3}{l}{\textbf{Radial Velocity Offsets and Jitters}} \\[0.05cm]
RV off$_{\rm FEROS}$ [m/s] & 13684.6 & 13683.3$_{-5.6}^{+7.5}$ \\
RV off$_{\rm HARPS}$ [m/s] & 30.7 & 29.8$_{-5.7}^{+4.1}$ \\
RV off$_{\rm ESPR}$ [m/s] & 13684.3 & 13681.3$_{-4.5}^{+5.5}$ \\
RV jitt$_{\rm FEROS}$ [m/s] & 34.3 & 32.2$_{-4.0}^{+6.4}$ \\
RV jitt$_{\rm HARPS}$ [m/s] & 19.0 & 22.8$_{-3.7}^{+3.1}$ \\
RV jitt$_{\rm ESPR}$ [m/s] & 9.1 & 13.0$_{-2.8}^{+5.6}$ \\[0.2cm]

\multicolumn{3}{l}{\textbf{Orbital Parameters}} \\[0.05cm]
$K_b$ [m/s] & 2.26 & 2.22$_{-0.18}^{+0.20}$ \\
$P_b$ [d] & 11.5076 & 11.50752$_{-0.00152}^{+0.00174}$ \\
$e\sin\omega_b$ & 0.0497 & 0.0517$_{-0.0039}^{+0.0038}$ \\
$e\cos\omega_b$ & 0.0494 & 0.0464$_{-0.0044}^{+0.0041}$ \\
$\lambda_b$ [deg] & 83.9 & 84.26$_{-0.49}^{+0.49}$ \\
$i_b$ [deg] & 89.9 & 89.94$_{-0.44}^{+0.53}$ \\[0.2cm]

$K_c$ [m/s] & 28.9 & 28.66$_{-1.65}^{+1.23}$ \\
$P_c$ [d] & 22.5635 & 22.56441$_{-0.00239}^{+0.00270}$ \\
$e\sin\omega_c$ & 0.0049 & 0.0068$_{-0.0055}^{+0.0051}$ \\
$e\cos\omega_c$ & 0.0313 & 0.0268$_{-0.0061}^{+0.0061}$ \\
$\lambda_c$ [deg] & 322.1 & 322.66$_{-0.73}^{+0.69}$ \\
$i_c$ [deg] & 90.3 & 90.05$_{-0.52}^{+0.43}$ \\[0.2cm]

$K_d$ [m/s] & 40.8 & 42.51$_{-2.63}^{+2.30}$ \\
$P_d$ [d] & 48.9273 & 48.9244$_{-0.0054}^{+0.0048}$ \\
$e\sin\omega_d$ & $-0.055$ & $-0.0492_{-0.0120}^{+0.0121}$ \\
$e\cos\omega_d$ & 0.0621 & 0.0548$_{-0.0097}^{+0.0088}$ \\
$\lambda_d$ [deg] & 166.1 & 167.01$_{-1.11}^{+1.18}$ \\
$i_d$ [deg] & 89.96 & 89.93$_{-0.44}^{+0.53}$ \\[0.2cm]

\multicolumn{3}{l}{\textbf{Derived Planetary and Orbital Quantities}} \\[0.05cm]
$m_p(b)$ [M$_{\rm J}$] & 0.0250 & 0.0250$_{-0.0022}^{+0.0024}$ \\
$m_p(c)$ [M$_{\rm J}$] & 0.4007 & 0.4033$_{-0.0251}^{+0.0237}$ \\
$m_p(d)$ [M$_{\rm J}$] & 0.7297 & 0.7728$_{-0.0520}^{+0.0514}$ \\
$a_b$ [AU] & 0.0996 & 0.10055$_{-0.00166}^{+0.00161}$ \\
$a_c$ [AU] & 0.1560 & 0.15754$_{-0.00260}^{+0.00253}$ \\
$a_d$ [AU] & 0.2615 & 0.26397$_{-0.00436}^{+0.00423}$ \\[0.2cm]
$R_b$ [R$_{\rm J}$] &  & $0.655\pm0.018$ \\
$R_c$ [R$_{\rm J}$] &  & $0.973\pm0.023$ \\
$R_d$ [R$_{\rm J}$] &  & $0.923\pm0.044$ \\[0.2cm]
$T_{eq,b}$ [K] &  & $872\pm8$ \\
$T_{eq,c}$ [K] &  & $696\pm7$ \\
$T_{eq,d}$ [K] &  & $538\pm5$ \\[0.2cm]
\hline
\end{tabular}
\label{tab:ttv_rv_posterior}
\tablecomments{The orbital elements are valid for epoch BJD = 2458656.3.}
\end{table}

\section{Discussion and Conclusions} \label{sec:concl}
In this work, we reported the discovery and characterisation of an exoplanetary system consisting of three transiting warm giant planets orbiting the solar-type G-dwarf star \stname, in a compact and nearly-resonant configuration. \plnameb\ is an inflated Neptune with an orbital period of \rPb. \plnamec\ is warm Saturn with an orbital period of \rPc. Finally, \plnamed\ is a warm Jupiter with an orbital period of \rPd. The low mutual inclinations of these three planets, their nearly-resonant configuration, and the low inferred stellar obliquity indicate that the planets probably experienced gentle disc migration from beyond the snowline \citep{Goldreich1979}.

\begin{figure}[t!]
    \centering
    \includegraphics[width=\columnwidth]{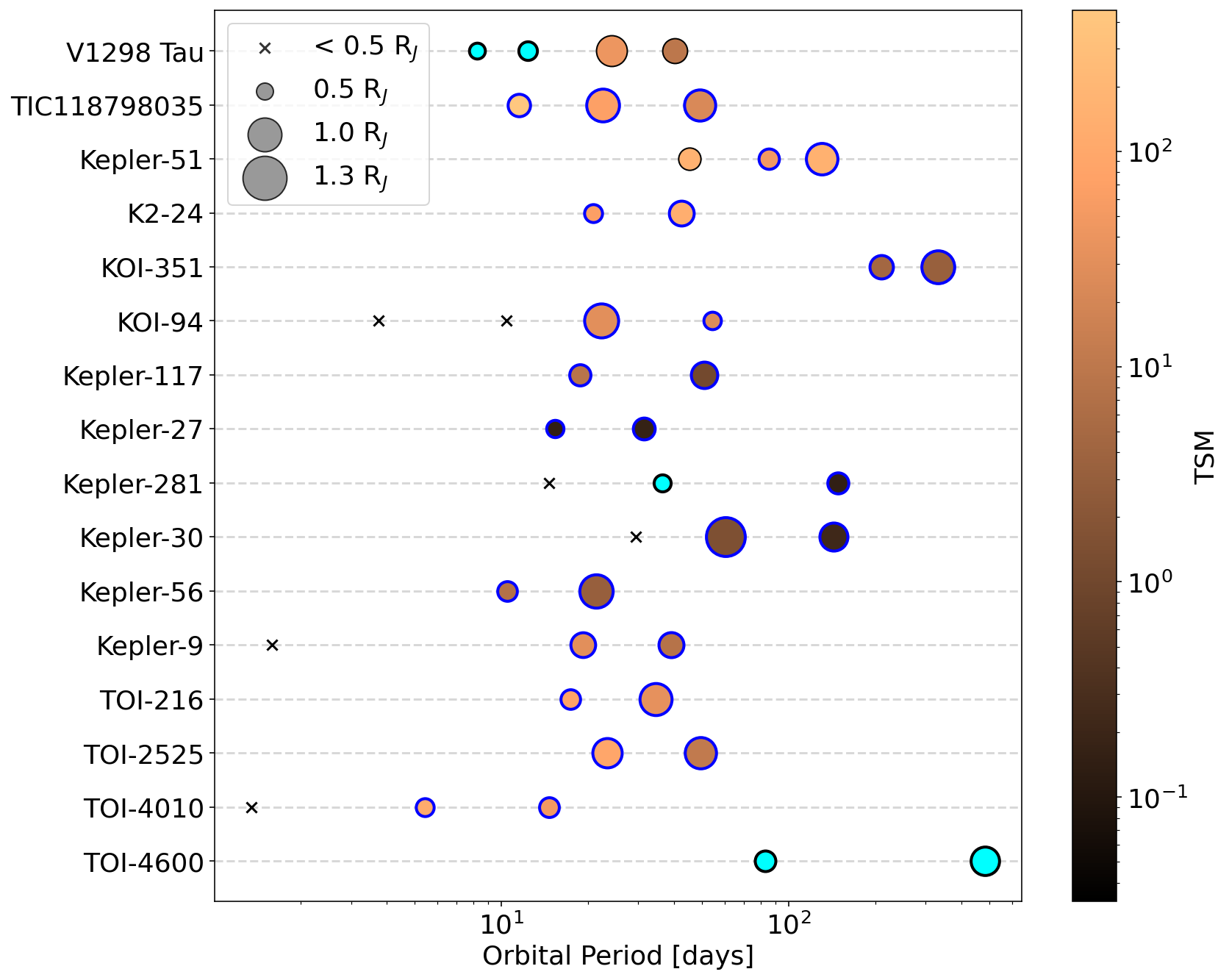}
    \caption{Systems containing two or more transiting planets with radius larger than 0.5 \rjup. The size of the circles scales up with the planet radius. Those planets with masses measured with a precision better than 20\% have blue borders. The circles are color-coded by TSM.}
    \label{fig:multis}
\end{figure}

The \stname\ system is one of the very few known systems that contains more than two giant ($R_P>0.5R_J$) transiting planets (see Figure \ref{fig:multis}). For example, the Kepler-31 system \citep{kepler31} presents three transiting giant planets, but all with radii similar to Neptune. The Kepler-51 system contains three planets with radii between Neptune and Jupiter, but all these planets fall in the rare category of super-puffs \citep{kepler-51}. Another comparable system is V1298 Tau \citep{david2019}, which has four transiting planets larger than Neptune, but due to the young age of the star, the masses of these planets are mostly unconstrained \citep{finociety2023}.
\stname\ stands out as the only system with three transiting giant planets with precise masses suitable for atmospheric studies.

\begin{figure}[t!]
    \centering
    \includegraphics[width=\columnwidth]{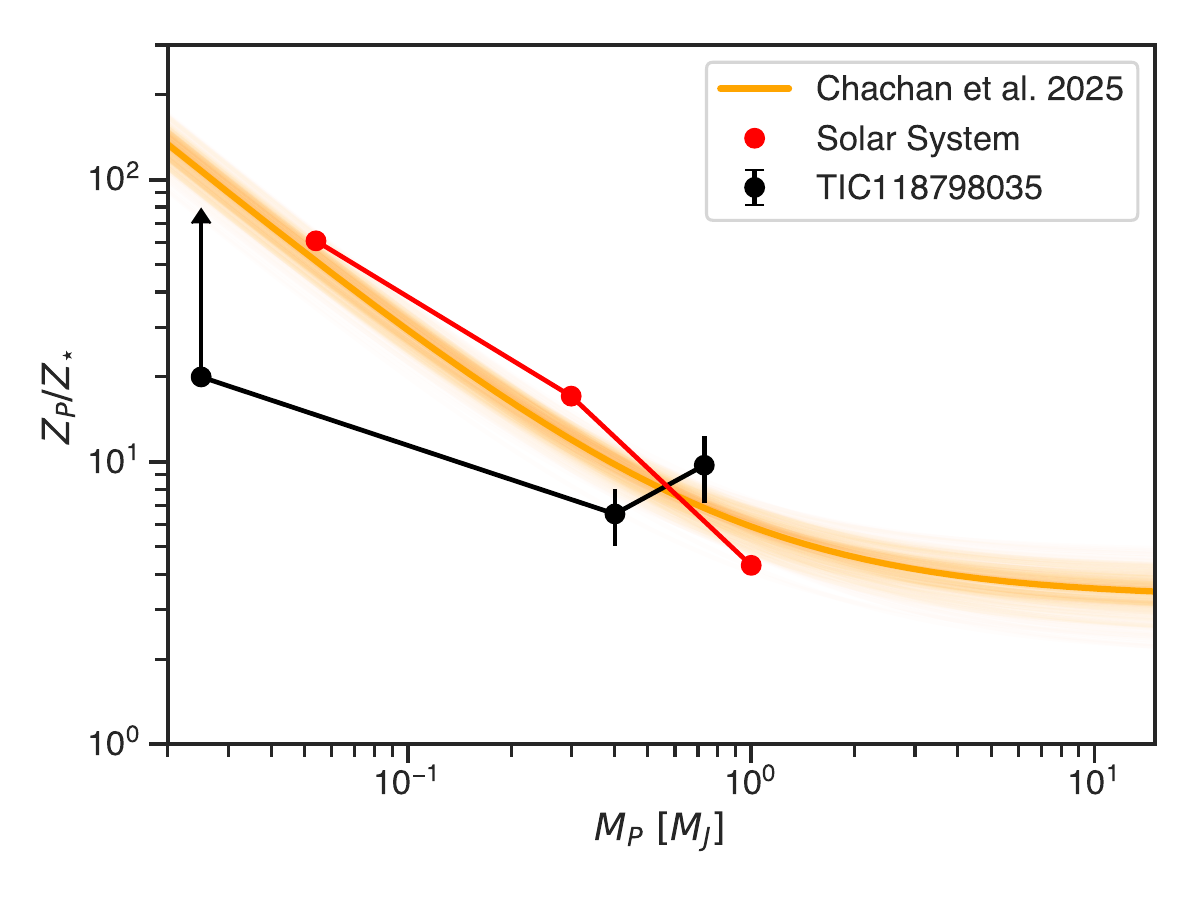}
    \caption{Relative planet bulk metallicities as a function of the planet mass for the \stname\ system (black circles), and the solar system (Neptune, Uranus, Jupiter, red circles). We also show the empirical relation found by \citep{chachan}.}
    \label{fig:zpzs}
\end{figure}

Due to the detailed characterization of the planets of the \stname\ system, we were able to model the interior structure of the planets and to infer their bulk metallicities. Figure \ref{fig:zpzs} shows the relative metal enrichment of these planets as a function of the planet mass, where we also indicate the values for three planets of the solar system (Neptune, Saturn, and Jupiter). The Neptune-mass planet in both systems seems to be significantly enhanced in metals compared to the gas giants. This is expected for core dominated planets that did not enter into the run-away gas accretion stage. However, we can identify an important difference between the \stname\ system and the solar system. The metallicities of \plnamec\ and d are comparable, and \plnamed\ could even have a higher fraction of heavy elements than \plnamec. This goes in the opposite direction of the mass-metallicity correlation seen in our solar system, but can still be consistent within the errorbars with the empirical mass-metallicity relation for warm giant planets found in \citet{chachan}. The total inferred amount of heavy elements in Saturn and Jupiter is comparable ($\sim$20 $M_{\oplus}$), indicating a roughly similar history of envelope enrichment with planetesimal/pebble throughout their evolution.
For the \stname\ system we find that the total amount of heavy elements of \plnamed\ ($55\pm12 M_{\oplus}$) is significantly larger than that of \plnamec\ ($18\pm4 M_{\oplus}$), which is consistent with the empirical correlation found by \citet{thorngren:2016}. This difference in the amount of solids accreted could be explained by differences in the migration distance covered by both planets \citep{shibata2020}, where \plnamed\ formed significantly beyond the snowline and was able to accrete a large fraction of planetesimals and/or pebbles \citep{danti2023} during its viscous migration into the inner parts of the system. In this case we would expect different atmospheric compositions for both exterior giants of the \stname\ system. Another possibility is that similar super-stellar bulk metallicities of both gas giants were produced by accretion of similar volatile-enriched gas due to the evaporation of icy pebbles that drifted from beyond the snowline \citep{Schneider:2021}. In this case, while accreting gas of similar composition inside the snowline, the difference in final masses for both planets could originate from a delayed runaway gas accretion of \plnamec\ compared to \plnamed\ \citep{Bergez-Casalou}. In this scenario we should expect similar atmospheric metallicities and C/O ratios for \plnamec\ and \plnamed, which could be tested in the future.

\plnameb\ is significantly larger than Neptune, but has half of its mass, and can therefore be categorized as a fluffy sub-Neptune mass planet in the ``Savanna" region \citep{bourrier2023}. \plnameb\ is among the sub-Neptune mass planets with the lowest bulk density. The mild insolation that it receives could be inflating the planet envelope, but given that this planet is outside the Neptunian ``Desert" and ``Ridge" \citep{ridge} it is able to securely retain its atmosphere from evaporation \citep{owen2019}. \plnameb\ was never massive enough to trigger the runaway gas accretion to become a jovian planet. One possibility is that the core of this planet was formed inefficiently inside the snowline from refractory material. Another possibility is that it was initially formed beyond the snowline \citep{Venturini2017} but the presence of the other two massive planets acted as a barrier for the inward flux of volatile pebbles and/or planetesimals, which prevented the core of \plnameb\ to become the embryo of a Jovian planet \citep{Bitsch2023}.

The \stname\ system is a unique target for future exoplanet atmospheric studies. With three transiting exoplanets spanning masses from Jupiter to sub-Neptunes, the system offers an opportunity to test mass-metallicity relationships that have been elusive to test with other multiplanet systems in the Neptune/sub-Neptune regime alone \citep[see, e.g.,][]{lb:2020, barat:2024}. As showcased in Figure \ref{fig:zpzs} the maximum atmospheric metallicities all range on numbers at or above $\times 10$ solar, implying well-known atmospheric features such as CO$_2$ on the more massive planets c and d should be detectable with instrumentation such as that onboard the \textit{JWST} \citep{co2}. The sub-Neptune mass planet TIC11879803~b, on the other hand, falls in an interesting region of parameter space --- it orbits a similar star and has very similar bulk properties to (although it is more inflated than) TOI-421~b, which has been recently shown to be a likely haze free, low metallicity world \citep{davenport:2025}. If TIC11879803~b has similar atmospheric properties to TOI-421~b, then its atmospheric features should be readily detectable with \textit{JWST}. Having a similar stellar brigthness than WASP-39, the planets in the TIC11879803 system might all be targeted with JWST's NIRSpec/PRISM with only mild saturation in the 1-2 $\mu$m region, allowing to capture H$_2$O and CO$_2$ features that are to be expected in the planets on the system \citep{rustamkulov:2023, carter:2024}. This atmospheric exploration would allow for a test of whether the mass-metallicity relationship observed in the atmospheres of Solar System planets also maps to systems elsewhere.

\begin{acknowledgments}
R.B. acknowledges support from FONDECYT Project 1241963 and from ANID -- Millennium  Science  Initiative -- ICN12\_009.
T.T. acknowledges support by the BNSF program "VIHREN-2021" project No. KP-06-DV/5.
A.J. acknowledges support from Fondecyt project 1251439.
This paper was based on observations collected at the European Southern Observatory under ESO programmes 114.27CV, 114.27CS, and 115.286G.

This paper also made use of data collected by the TESS mission and are publicly available from the Mikulski Archive for Space Telescopes (MAST) operated by the Space Telescope Science Institute (STScI). Funding for the TESS mission is provided by NASA's Science Mission Directorate. We acknowledge the use of public TESS data from pipelines at the TESS Science Office and at the TESS Science Processing Operations Center. Resources supporting this work were provided by the NASA High-End Computing (HEC) Program through the NASA Advanced Supercomputing (NAS) Division at Ames Research Center for the production of the SPOC data products.

This research has made use of the Exoplanet Follow-up Observation Program (ExoFOP) and NASA Exoplanet Archive websites, which are operated by the California Institute of Technology under contract with NASA under the Exoplanet Exploration Program. 

Based on observations obtained with the HATPI project at the Las Campanas Observatory of the Carnegie Institution for Science. HATPI is operated by the Department of Astrophysical Sciences at Princeton
University. External funding for HATPI has been provided by the Gordon and Betty Moore Foundation, the David and Lucile Packard Foundation, the Mount Cuba Astronomical Foundation, and the Agencia Nacional de
Investigación y Desarrollo (ANID) of Chile through QUIMAL, Millennium
and Fondecyt grants.

\end{acknowledgments}





%
\facilities{VLT/ESPRESSO, TESS, ESO:3.6m (HARPS), Max Planck:2.2m (FEROS), HATPI, Observatoire Moana.}

\software{
\texttt{astropy} \citep{astropy},
\texttt{batman} \citep{batman},
\texttt{ceres} \citep{ceres},
\texttt{celerite} \citep{celerite},
\texttt{dynesty} \citep{dynesty},
\texttt{exostriker} \citep{exostriker},
\texttt{juliet} \citep{juliet},
\texttt{lightkurve} \citep{lightkurve}
\texttt{radvel} \citep{radvel},
\texttt{serval} \citep{serval},
\texttt{zaspe} \citep{zaspe}.
          }

\appendix

\begin{table*}
\begin{center}
\caption{Radial velocity and activity indices of \stname.}
\label{tab:rvs}
\scalebox{0.9}{
\begin{tabular}{cccccc}

\hline \hline
BJD             & RV [m/s]           & Bisector Span    & FWHM    & S-index & Instrument\\ \hline 
\hline \hline
2460516.66457690 &	$13617.1 \pm   9.8$	&    $-6.0 \pm   13.0$	&  $10276.9 \pm   13.0$		&  $-0.1811 \pm     0.0191$	&	FEROS	\\
2460518.68752548 &	$13617.3 \pm  11.0$	&    $24.0 \pm   15.0$	&  $10185.7 \pm   15.0$		&  $0.2499 \pm      0.0230$	&	FEROS	\\
2460534.63568115 &	$13726.7 \pm   9.5$	&    $-5.0 \pm   13.0$	&  $10172.0 \pm   13.0$		&  $0.2315 \pm      0.0154$	&	FEROS	\\
2460536.57691066 &	$13696.6 \pm   9.5$	&    $-9.0 \pm   13.0$	&  $10194.3 \pm   13.0$		&  $0.2106 \pm      0.0167$	&	FEROS	\\
2460537.56606724 &	$13711.8 \pm   9.8$	&    $47.0 \pm   13.0$	&  $10354.1 \pm   13.0$		&  $0.2163 \pm      0.0177$	&	FEROS	\\
2460544.63757318 &	$13665.9 \pm  13.1$	&  $-149.0 \pm   16.0$	&  $10309.1 \pm   16.0$		&  $0.3405 \pm      0.0340$	&	FEROS	\\
2460545.52587118 &	$13743.9 \pm  12.4$	&    $24.0 \pm   16.0$	&  $10144.2 \pm   16.0$		&  $0.1747 \pm      0.0287$	&	FEROS	\\
2460546.56269445 &	$13649.8 \pm  11.7$	&    $-1.0 \pm   15.0$	&  $10114.8 \pm   15.0$		&  $0.1928 \pm      0.0228$	&	FEROS	\\
2460547.51458969 &	$13699.3 \pm  10.0$	&   $-45.0 \pm   13.0$	&  $10259.3 \pm   13.0$		&  $0.1933 \pm      0.0180$	&	FEROS	\\
2460582.56663757 &	$13751.2 \pm  10.8$	&   $-13.0 \pm   14.0$	&  $10489.6 \pm   14.0$		&  $0.3175 \pm      0.0298$	&	FEROS	\\
2460584.63506224 &	$13669.7 \pm  14.8$	&   $147.0 \pm   18.0$	&  $10312.7 \pm   18.0$		&  $-0.0655 \pm      0.0577$	&	FEROS	\\
2460589.59159943 &	$13718.5 \pm   3.2$	&  $-34.1 \pm    5.0$	&   $7652.8 \pm    5.0$		&  $-0.015 \pm      0.009$	&	ESPRESSO	\\
2460595.53553033 &	$13720.6 \pm   5.4$	&   $-35.1 \pm    6.0$	&   $7653.8 \pm    6.0$		&  $-0.136 \pm      0.014$	&	ESPRESSO	\\
2460605.53406674 &	$13665.7 \pm   3.1$	&   $-73.7 \pm   14.0$	&   $7702.4 \pm   14.0$		&  $-0.839 \pm      0.050$	&	ESPRESSO	\\
2460605.54883009 &	  $-10.0 \pm   6.2$	&   $-36.0 \pm   16.0$	&   $8446.4 \pm   16.0$		&  $-1.5276 \pm      0.0930$	&	HARPS	\\
2460610.55675713 &	  $-58.0 \pm   4.1$	&    $-9.0 \pm   11.0$	&   $8537.3 \pm   11.0$		&  $-0.7456 \pm      0.0466$	&	HARPS	\\
2460622.52120867 &	$13684.4 \pm  14.3$	&   $-44.0 \pm   17.0$	&  $10537.8 \pm   17.0$		&  $0.1895 \pm      0.0407$	&	FEROS	\\
2460625.51130979 &	$13665.5 \pm  10.3$	&    $52.0 \pm   13.0$	&  $10685.5 \pm   13.0$		&  $0.2383 \pm      0.0420$	&	FEROS	\\
2460741.87221931 &	   $21.8 \pm   9.4$ 	&   $-14.0 \pm   23.0$	&   $8600.8 \pm   23.0$		&  $-3.6925 \pm      0.2544$	&	HARPS	\\
2460745.84740885 &	   $-4.5 \pm   2.3$	&    $-3.0 \pm    7.0$	&   $8486.4 \pm    7.0$		&  $0.0034 \pm      0.0097$	&	HARPS	\\
2460735.85841482 &	$13750.1 \pm   2.7$	&  $-43.1 \pm    8.0$	&   $7643.1 \pm    8.0$		&  $-0.3460 \pm      0.020$	&	ESPRESSO	\\
2460738.80407461 &	$13704.0 \pm   7.1$	&   $-12.7 \pm   10.0$	&   $7700.6 \pm   10.0$		&  $-0.358 \pm      0.027$	&	ESPRESSO	\\
2460740.88443271 &	$13725.2 \pm   4.4$	&   $-63.5 \pm    5.0$	&   $7649.4 \pm    5.0$		&  $-0.021 \pm      0.009$	&	ESPRESSO	\\
2460743.84103942 &	$13678.7 \pm   2.7$	&   $-42.9 \pm    6.0$	&   $7623.0 \pm    6.0$		&  $-0.061 \pm      0.011$	&	ESPRESSO	\\
2460762.87223892 &	   $16.8 \pm   9.5$	&    $12.0 \pm   12.0$	&   $8551.8 \pm   12.0$		&  $-0.0571 \pm      0.0215$	&	HARPS	\\
2460763.89050326 &	    $1.0 \pm   5.7$	&    $-6.0 \pm    7.0$	&   $8499.0 \pm    7.0$		&  $-0.0148 \pm      0.0096$	&	HARPS	\\
2460775.88615376 &	   $85.4 \pm   4.9$	&   $-30.0 \pm   15.0$	&   $8483.7 \pm   15.0$		&  $-0.4836 \pm      0.0311$	&	HARPS	\\
2460790.90241155 &	   $-1.2 \pm   1.6$	&    $17.0 \pm    4.0$	&   $8480.7 \pm    4.0$		&  $0.1162 \pm      0.0075$	&	HARPS	\\
2460805.91564241 &	   $24.1 \pm   3.9$	&    $10.0 \pm   12.0$	&   $8474.1 \pm   12.0$		&  $0.0412 \pm      0.0080$	&	HARPS	\\
2460808.68981745 &	$13823.0 \pm  13.8$	&    $41.0 \pm   17.0$	&  $10687.5 \pm   17.0$		&  $0.4568 \pm      0.0448$	&	FEROS	\\
2460819.81270876 &	   $36.7 \pm   4.9$	&   $-13.0 \pm    6.0$	&   $8508.7 \pm    6.0$		&  $0.0569 \pm      0.0097$	&	HARPS	\\
2460832.81446173 &	$13650.4 \pm   9.2$	&    $19.0 \pm   12.0$	&  $10145.5 \pm   12.0$		&  $0.1240 \pm      0.0145$	&	FEROS	\\
2460833.90964164 &	   $19.4 \pm   3.3$	&     $5.0 \pm   11.0$	&   $8521.9 \pm   11.0$		&  $-0.3220 \pm      0.0218$	&	HARPS	\\
2460860.69465978 &	  $-38.5 \pm   2.0$	&    $19.0 \pm    6.0$	&   $8485.1 \pm    6.0$		&  $0.0415 \pm      0.0090$	&	HARPS	\\
2460868.68595354 &	   $44.2 \pm   3.2$	&    $20.0 \pm   10.0$	&   $84503.0 \pm  10.0$		&  $-0.0605 \pm      0.0171$	&	HARPS	\\
2460872.67188090 &	   $50.7 \pm   1.9$	&     $3.0 \pm    5.0$	&   $8486.6 \pm    5.0$		&  $0.0686 \pm      0.0084$	&	HARPS	\\
2460875.75508209 &	$13706.3 \pm  10.6$	&   $-57.0 \pm   14.0$	&  $10150.9 \pm   14.0$		&  $0.2098 \pm      0.0219$	&	FEROS	\\
2460900.63233013 &	$13538.4 \pm  10.6$	&   $-37.0 \pm   14.0$	&  $10224.8 \pm   14.0$		&  $0.1580 \pm      0.0211$	&	FEROS	\\
2460911.61214084 &	$13657.7 \pm  10.2$	&   $-17.0 \pm   13.0$	&  $10451.7 \pm   13.0$		&  $0.3306 \pm      0.0211$	&	FEROS	\\
2460912.60567926 &	   $49.5 \pm   2.0$	&    $13.0 \pm    6.0$	&   $8530.9 \pm    6.0$		&  $0.0421 \pm      0.0103$	&	HARPS	\\
2460913.59478035 &	$13764.6 \pm   9.4$	&   $-34.0 \pm   13.0$	&  $10220.5 \pm   13.0$		&  $0.2519 \pm      0.0166$	&	FEROS	\\
2460927.66625271 &	$13727.6 \pm  12.2$	&     $5.0 \pm   15.0$	&  $10237.7 \pm   15.0$		&  $0.3896 \pm      0.0389$	&	FEROS	\\
2460928.58258090 &	$13752.4 \pm  12.5$	&   $-50.0 \pm   15.0$	&  $10451.5 \pm   15.0$		&  $0.1957 \pm      0.0269$	&	FEROS	\\
2460928.63713579 &	   $59.6 \pm   2.5$	&     $3.0 \pm    7.0$	&   $8532.0 \pm    7.0$		&  $-0.0232 \pm      0.0141$	&	HARPS	\\
2460929.61823672 &	$13674.3 \pm  13.1$	&    $11.0 \pm   16.0$	&  $10334.0 \pm   16.0$		&  $0.3500 \pm      0.0274$	&	FEROS	\\
2460930.49609218 &	$13714.8 \pm   8.2$	&    $42.0 \pm   11.0$	&  $10270.0 \pm   11.0$		&  $0.2717 \pm      0.0131$	&	FEROS	\\
2460938.65897170 &	$13812.9 \pm  15.0$	&  $-106.0 \pm   18.0$	&  $10038.6 \pm   18.0$		&  $0.2688 \pm      0.0720$	&	FEROS	\\
2460940.58919492 &	$13741.9 \pm  15.3$	&   $-95.0 \pm   18.0$	&  $10568.6 \pm   18.0$		&  $0.0732 \pm      0.0421$	&	FEROS	\\
2460941.61598775 &	$13726.8 \pm  13.7$	&    $32.0 \pm   17.0$	&  $10076.6 \pm   17.0$		&  $0.3249 \pm      0.0451$	&	FEROS	\\
2460943.58286111 &	$13708.2 \pm  14.8$	&   $-62.0 \pm   18.0$	&  $10164.3 \pm   18.0$		&  $0.3884 \pm      0.0470$	&	FEROS	\\
\end{tabular}}
\end{center}
\end{table*}


\bibliography{sample701}{}
\bibliographystyle{aasjournalv7}



\appendix
\setcounter{table}{0}
\renewcommand{\thetable}{A\arabic{table}}

\setcounter{figure}{0}
\renewcommand{\thefigure}{A\arabic{figure}}

\begin{figure*}[tp]
\begin{center}$
\begin{array}{ccc} 
	\includegraphics[width=17.50cm]{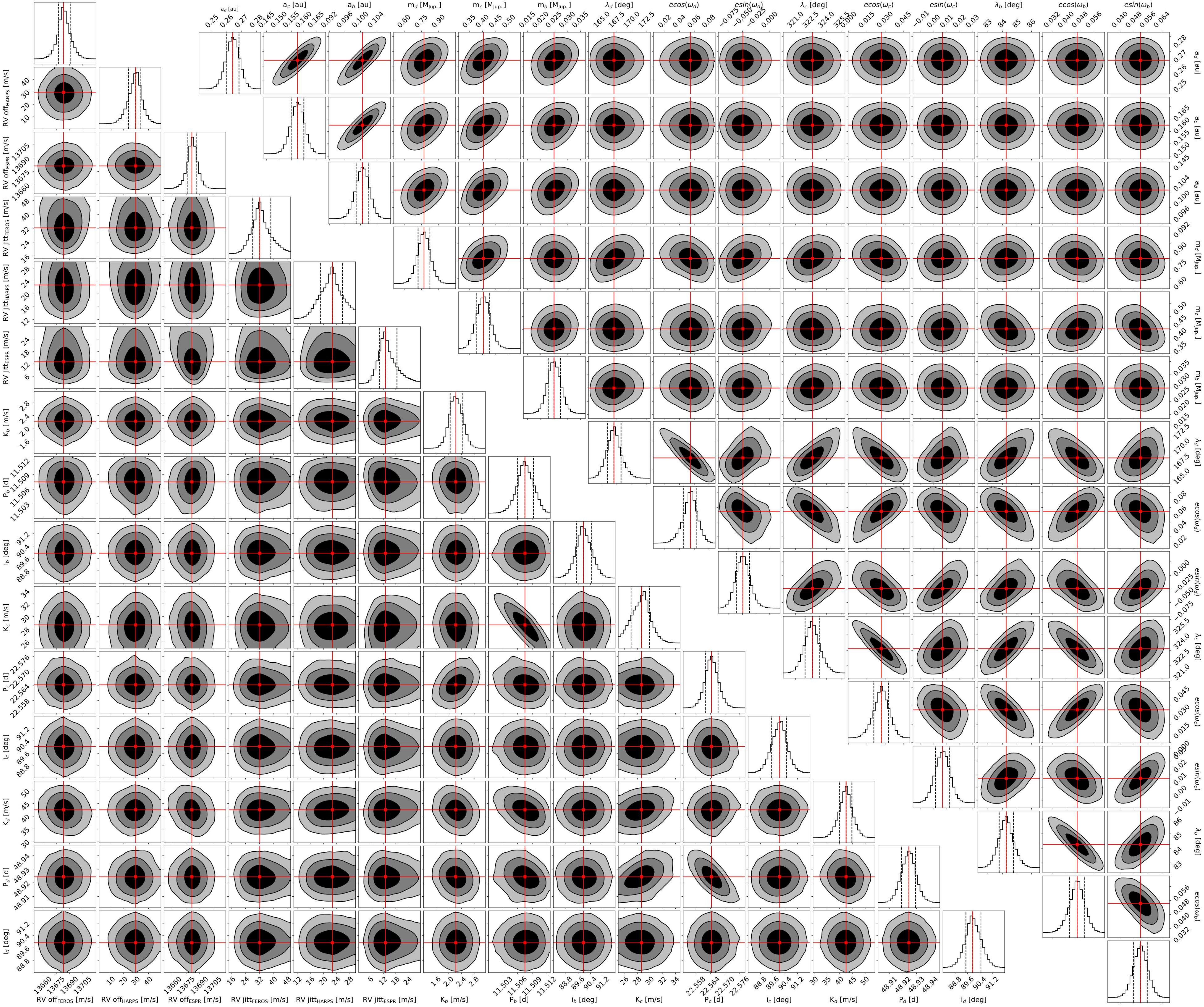} 
\end{array} $
\end{center}
\caption{Posterior probability distributions from the joint TTV and RV MCMC modeling of TIC\,118798035. 
The median values of the fitted and derived parameters are indicated by red crosses. 
Black contours mark the 1, 2, and 3$\sigma$ confidence intervals of the distributions.}

\label{mcmc_samp} 
\end{figure*}

\end{document}